\documentclass[preprint]{aastex}

\newcommand\lta{\mathrel{\hbox{\raise 0.6 ex \hbox{$<$}\kern
                   -1.8 ex\lower .5 ex\hbox{$\sim$}}}}
\newcommand\gta{\mathrel{\hbox{\raise 0.6 ex \hbox{$>$}\kern
                   -1.7 ex\lower .5 ex\hbox{$\sim$}}}}
 
\shortauthors{VandenBerg et al.}
\shorttitle{Stellar Models with Convective Overshooting}
 
\begin{document}
 
\title{The Victoria--Regina Stellar Models: Evolutionary Tracks and Isochrones
for a Wide Range in Mass and Metallicity that Allow for Empirically Constrained
Amounts of Convective Core Overshooting}
 
\author{Don A.~VandenBerg}
\affil{Department of Physics \& Astronomy, University of Victoria,
       P.O.~Box 3055, Victoria, B.C., V8W~3P6, Canada}
\email{davb@uvvm.uvic.ca}

\author{Peter A.~Bergbusch}
\affil{Department of Physics, University of Regina, Regina,
       Saskatchewan, S4S~0A2, Canada}
\email{bergbush@phys.uregina.ca}
 
\author{Patrick D.~Dowler}
\affil{Herzberg Institute of Astrophysics, National Research Council Canada,
       5071 W.~Saanich Rd., Victoria, B.C., V9E~2E7, Canada}
\email{Patrick.Dowler@nrc-cnrc.gc.ca}
 
\begin{abstract}
Seventy-two grids of stellar evolutionary tracks, along with the capability
to generate isochrones and luminosity/color functions from them, are presented
in this investigation.\footnote{All of the model grids may be obtained from
the Canadian Astronomy Data Center (http://www.cadc-ccda.hia-
iha.nrc-cnrc.gc.ca/cvo/community/VictoriaReginaModels/).  Included in this
archive are (i) the interpolation software (FORTRAN 77) to produce isochrones,
isochrone probability functions, luminosity functions, and color functions,
along with instructions on how to implement and use the software, (ii) $BVRI$
(VandenBerg \& Clem 2003) and $uvby$ (\citealt{cvg04}) color-temperature
relations, and (iii) Zero-Age Horizontal Branch loci for all of the chemical
compositions considered.}   Sixty of them extend (and encompass) the sets of
models reported by VandenBerg et al.~(2000, ApJ, 532, 430) for 17 [Fe/H] values
from $-2.31$ to $-0.30$ and $\alpha$-element abundances corresponding to
[$\alpha$/Fe] $=0.0, 0.3,$ and 0.6 (at each iron abundance) to the solar
metallicity and to sufficiently high masses (up to $\sim 2.2 {{\cal M}_\odot}$)
that isochrones may be computed for ages as low as 1 Gyr.  The remaining grids
contain tracks for masses from 0.4 to 4.0 ${{\cal M}_\odot}$ and 12 [Fe/H]
values between $-0.60$ and $+0.49$ (assuming solar metal-to-hydrogen number
abundance ratios): in this case, isochrones may be calculated down to $\sim 0.2$
Gyr.  The extent of convective core overshooting has been modelled using a
parameterized version of the Roxburgh (1989, A\&A, 211, 361) criterion, in which
the value of the free parameter at a given mass and its dependence on mass have
been determined from analyses of binary star data and the observed
color-magnitude diagrams for several open clusters.  Because the calculations
reported herein satisfy many empirical constraints, they should provide useful
probes into the properties of both simple and complex stellar populations.

\end{abstract}
 
 
\section{Introduction}
\label{sec:intro}
It has been known for well over a decade that stellar models for
intermediate-mass and massive stars must allow for some degree of convective
core overshooting (CCO) if they are to provide satisfactory representations of
observed stars --- see, e.g., \citet{cbb92} for a brief review of the landmark
papers since the early 1970s that have contributed to this result and
\citet{chi99} for a recent summary of the consequences of CCO for stellar
evolution.  Despite the considerable progress that has been made in developing
a theory for turbulent convection (e.g., \citealt{rox89}; \citealt{xcd97};
\citealt{cd98}; \citealt{can99}; and references mentioned therein), it is still
not possible to predict the sizes of convective cores in stars from first
principles.  This is due largely to the difficulty of predicting the rate of
dissipation of turbulent kinetic energy.  As a result, there has been little
recourse but to use observations of, in particular, eclipsing binary stars
(e.g., \citealt{and91}; \citealt{spe97}; \citealt{rjg00}; \citealt{grf00}) and
open clusters (e.g., \citealt{dlm94}; \citealt{kdp97}; \citealt{rv98}) to
constrain the amounts of CCO that are assumed in the model computations.  

Based on such empirical considerations, most of the large grids of evolutionary
tracks and isochrones currently in use (e.g., those by \citealt{mms94};
\citealt{gbb00}; \citealt{ydk01}) have assumed that convective cores are
enlarged by the equivalent of $\sim 0.2 H_P$, where $H_P$ is the pressure scale
height at the classical boundary of the convective core (i.e., as given by the
Schwarzschild criterion).  This enlargement is taken to be independent of mass,
except between $\approx 1.5$ and $1.1 {{\cal M}_\odot}$, where the amount of
overshooting is generally assumed to decrease from $\sim 0.2$ to $0.0 H_P$.
Efforts are being made to determine whether this calibration, which has been
inferred primarily from observations of binary stars and star clusters having
close to the solar metallicity, applies to metal-poor, intermediate-age stars
like those found in stellar systems belonging to the Large Magellanic Cloud (see
\citealt{kdb01}; \citealt{bng03}; \citealt{wgd03}).  Although the uncertainties
are still too large to preclude some dependence of the overshooting distance on
metal abundance, models for metal-deficient stars that allow for $\sim 0.2 H_P$
extensions of the convective core appear to fare quite well in explaining the
color-magnitude diagrams (CMDs) of LMC clusters (also see \citealt{rsz01}).

In the present study, we have opted to use a physics-based criterion to predict
the sizes of convective cores.  By integrating the full equations of
compressible fluid dynamics, under reasonable assumptions (see \citealt{zah91}),
\citet{rox89} derived the following integral equation for the maximum size of a
central convective zone, $r_{\rm max}$ (in the case that viscous dissipation is
negligible):
\begin{equation}
\int\limits_0^{r_{\rm max}}(L_{\rm rad}-L){1\over
   T^2}{{d\,T}\over{d\,r}}d\,r = 0~.
\label{eq:rox}
\end{equation}
Here, $L_{\rm rad}$ and $L$ represent, in turn, the radiative luminosity and
the total luminosity produced by nuclear reactions, while the other symbols
have their usual meanings.  If the viscous dissipation of energy is taken into
account, then (as described by Roxburgh) a second integral is introduced and it
is the radius where the difference between the two integrals vanishes that
defines the {\it actual} boundary of a convective core, $r_{cc}$. Unfortunately,
it is not yet possible to evaluate the dissipation term; consequently, we have
chosen to represent the second integral as a fraction of the first integral and
to derive $r_{cc}$ from
\begin{equation}
\int\limits_0^{r_0}F_{\rm over}(L_{\rm rad}-L){1\over
 T^2}{{d\,T}\over{d\,r}}d\,r + \int\limits_{r_0}^{r_{\rm cc}}(2-F_{\rm
 over})(L_{\rm rad}-L){1\over T^2}{{d\,T}\over{d\,r}}d\,r = 0~.
\label{eq:rox1}
\end{equation}
This involves the free parameter $F_{\rm over}$, which must be calibrated using
observations.  [The integral constraint must be written this way because
$L_{\rm rad} - L$ changes sign at the classical core boundary, $r_0$, whereas
the dissipation term is always positive.  Note, as well, that bigger values of
$F_{\rm over}$ imply larger overshooting zones: it is apparent, for instance,
that equation~(\ref{eq:rox}) is recovered when $F_{\rm over} = 1$.]  Granted,
this approach is still quite {\it ad hoc}, but at least the evaluation of
$1 - F_{\rm over}$ provides a measure of the relative importance of the
dissipation term in stars of different mass and chemical abundances --- which
may ultimately help to constrain convection theory. 

Sufficient work has been carried out to date (by \citealt{dow94};
\citealt{rv98}; \citealt{gvs98}; \citealt{vs04}; and \citealt{dv05}) to
permit a calibration of $F_{\rm over}$ as a function of mass for stars
having [Fe/H] $\approx 0.0$.  This calibration is presented and
discussed in \S 2 along with a suggestion of how that calibration might
differ as a function of metallicity.  The codes used to compute both
the evolutionary tracks and the isochrones (as well as isochrone
probability, luminosity, and color functions) are briefly described in
\S 3.  This section also contains a summary of the properties of the
models for which evolutionary tracks have been generated.  In \S 4 some
additional tests of the models are presented, while brief concluding
remarks are given in \S 5.
  
\section{The Dependence of $F_{\rm over}$ on Mass and Metallicity}
\label{sec:fover}
The open clusters M$\,$67, NGC$\,$6819, and NGC$\,$7789 provide a particularly
suitable set of clusters for the calibration of $F_{\rm over}$ as a function of
stellar mass because (i) they are populous systems with well-defined CMDs, (ii)
they span a fairly wide range in age (from $\approx 4$ Gyr to $\approx 1.7$ Gyr,
see below), and (iii) their metal abundances appear to be the same to within
$\delta\,$[Fe/H] $\approx 0.1$ dex.  As discussed by \citet{vs04}, the
metallicity of M$\,$67 seems to be especially well established at [Fe/H]
$= -0.04\pm 0.05$ (see, in particular, the high-resolution spectroscopic studies
by \citealt{ht91}; \citealt{tet00}).  The same value of [Fe/H] was recently
obtained for NGC$\,$7789 by \citet{tep05} using very similar techniques as in
their study of M$\,$67.  In the case of NGC$\,$6819, the solar (or a slightly
higher) metallicity is suggested by the latest findings (though earlier work
indicated a preference for [Fe/H] $\approx -0.1$; see the discussion by
\citealt{rv98}).  Based on their extensive moderate-resolution spectra (for many
open clusters), \citet{fjt02} concluded that NGC$\,$6819 is about 0.04 dex more
metal rich than M$\,$67, implying an [Fe/H] $\approx 0.0$.  An even higher
[Fe/H] value, by $\approx 0.1$ dex, was obtained by \citet{bcg01} from
high-dispersion spectra.

It is generally accepted that the amount of CCO in the turnoff (TO) stars of
M$\,$67 is significantly less than that which occurs in the upper main-sequence
(MS) stars of appreciably younger systems.\footnote{In fact, models that allow
for diffusive processes (gravitational settling and radiative accelerations)
are very successful in explaining the M$\,$67 CMD without invoking {\it any}
overshooting from convective cores (see \citealt{mrr04}).  However, the
enlargement of central convective cores due to such processes is small;
consequently, one can anticipate that it will not be possible for diffusive
models to avoid the need for CCO if they are to provide satisfactory
representations of the CMDs of younger open clusters.  Overshooting thus
provides an additional parameter that can be used to fine-tune the agreement
between synthetic and observed CMDs whether or not diffusive processes are
taken into account.  There is certainly little to choose between the
non-diffusive and diffusive isochrones that were fitted to M$\,$67 observations
by \citet{vs04} and by Michaud et al.  The main difference, as shown in the
latter study, is that the age at a given TO luminosity is reduced by 5--7\%
if diffusion is treated.}  For instance, \citet{svk99} have found that stellar
models which assume a $0.1 H_P$ enlargement of the convective core (over that
determined from the Schwarzschild criterion) provide a good match to the
M$\,$67 CMD, whereas this extension must be $\sim 0.2$--$0.25 H_P$ in order to
obtain comparable fits to the TO photometry of NGC$\,$752 and NGC$\,$3680 (see
\citealt{kdp97}).  Qualitatively similar results have been found using the
parameterized form of the Roxburgh criterion described in \S 1.  That is,
(non-diffusive) isochrones are able to reproduce the detailed morphology of
M$\,$67's turnoff only if they assume a much smaller value of $F_{\rm over}$
($\approx 0.07$; \citealt{vs04}) that that needed to fit the NGC$\,$6819
($\approx 0.5$; \citealt{rv98}) and NGC$\,$7789 ($\approx 0.5$; \citealt{gvs98})
CMDs.  As NGC$\,$6819 is only $\sim 1.5$ Gyr younger than M$\,$67, these results
imply that the extent of overshooting must increase very rapidly over a
relatively small range in TO mass.

Indeed, the values of $F_{\rm over}$ that were derived in the aforementioned
studies apply specifically to the most massive cluster stars that are still on
the main sequence.  For instance, the isochrones used by \citet{vs04} to fit
the M$\,$67 CMD predict that the stars which are just about to exhaust hydrogen
at their centers have masses of $\sim 1.25 {{\cal M}_\odot}$.  Only if $F_{\rm
over} \approx 0.07$ in stars of this mass is it possible for the models to
reproduce the observed turnoff morphology, including the luminosity of the gap
(see their Fig.~1).  Similarly, the value of $F_{\rm over} \approx 0.5$ derived
by \citet{rv98} from their comparison of solar metallicity isochrones to the
NGC$\,$6819 CMD is applicable to the brightest MS stars in this system:
according to the models, they have masses of $\approx 1.55 {{\cal M}_\odot}$.
By the same token, $F_{\rm over} \approx 0.5$ (or slightly larger, judging
from the plots provided by \citealt{gvs98}) must be adopted for $\sim 1.8
{{\cal M}_\odot}$ stars in order to achieve the good agreement between theory
and observations of NGC$\,$7789 reported by Gim et al.
    
\begin{figure}
\plotone{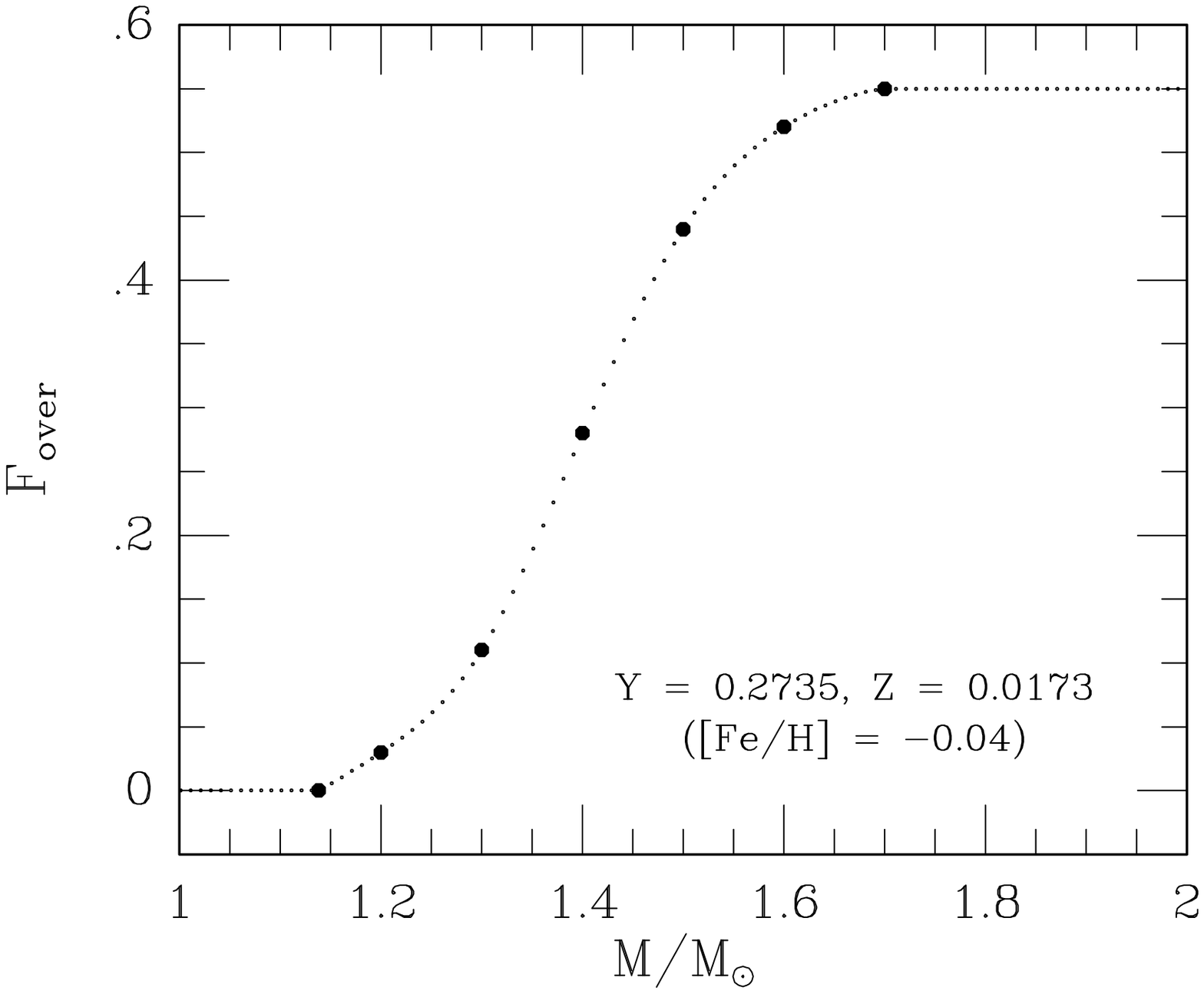}
\caption{The variation of $F_{\rm over}$ with mass that has been assumed in the
 computation of evolutionary tracks for stars having the indicated chemical
 abundances.}
\label{fig:fig1}
\end{figure}

These results suggest that something like the relationship between
$F_{\rm over}$ and mass shown in Figure~\ref{fig:fig1} applies to stars
having [Fe/H] $= -0.04$ (given that, as discussed above, two of the calibrating
clusters have this metal abundance, while the third one appears to be only
slightly more metal rich). At the moment, we have no basis for saying whether
$F_{\rm over}$ remains constant at masses $\gta 1.8 {{\cal M}_\odot}$ or
whether it slowly increases or decreases with increasing mass above $\approx
1.8 {{\cal M}_\odot}$, but we have opted to assume that the overshooting
parameter does not vary (i.e, $F_{\rm over} = 0.55$) in MS stars of higher mass
until observations tell us otherwise.  Non-overshooting models predict that
stars having [Fe/H] $= -0.04$ do not retain convective cores throughout their
MS lifetimes if they are less massive than $1.137 {{\cal M}_\odot}$.  Setting
$F_{\rm over} = 0.0$ at this, and lower, masses would seem to be a reasonable
assumption in view of the apparent steep dependence of $F_{\rm over}$ on mass
above $\approx 1.2 {{\cal M}_\odot}$.

\begin{figure}
\plotone{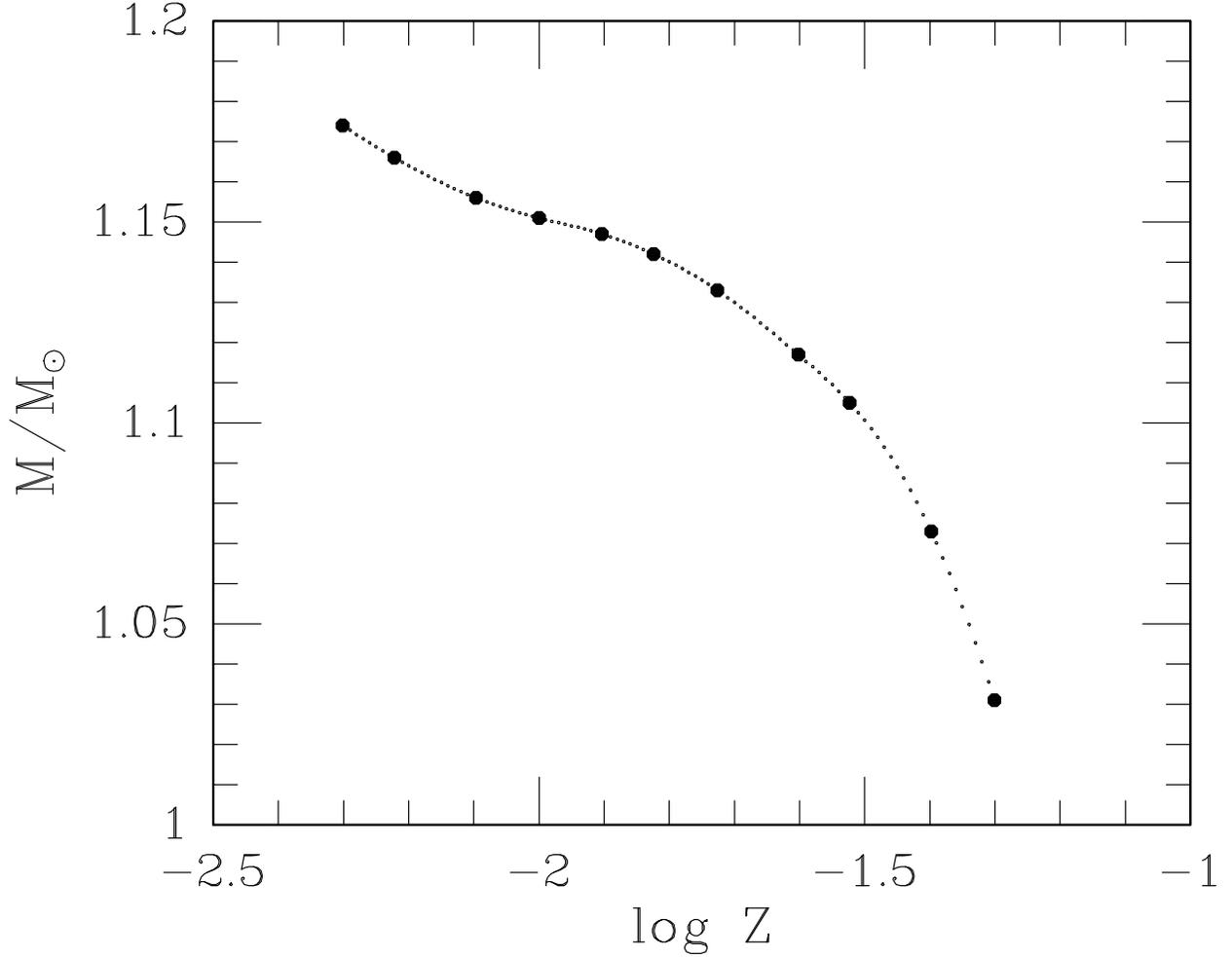}
\caption{Plot, as a function of $\log Z$, of the mass at which a transition
 is made between lower mass stars, on the one hand, and higher mass stars, on
 the other, that have radiative and convective cores, respectively, when they
 reach the end of the main-sequence phase.  These results are based on the
 ``VRSS" grids discussed in \S 3, which were computed for values of $Z$ in the
 range $0.005\le Z\le 0.05$.  (Although not shown, similar loci were computed
 for the sets of tracks having [Fe/H] values from $-2.31$ to 0.0 and
 [$\alpha$/Fe] $=0.0$, 0.3, and 0.6.)}  
\label{fig:fig2}
\end{figure}

This calibration of $F_{\rm over}$ probably does not apply to stars of different
metallicity because the transition from stars which have radiative centers at
the end of the MS phase to those which have convective cores until central H
exhaustion occurs at a mass that is a function of metal abundance. 
Figure~\ref{fig:fig2} illustrates how this transition mass varies with
$Z$ in the range $-2.3\le\log Z\le -1.3$ (i.e., from $Z=0.005$ to 0.05).  (These
results, which assume solar number abundance ratios of the heavy elements, were
derived from the ``VRSS" grids that are presented and discussed in the next
section.)  We see that, especially for super-metal-rich stars, the higher the
metallicity, the lower the mass at which convective cores persist throughout
the main-sequence phase.

Obviously, it would not make any sense to assume that there is no overshooting
in, e.g., a $Z=0.04$ star having a mass of $1.137 {{\cal M}_\odot}$ (the
limiting mass below which  $F_{\rm over} = 0.0$ if $Z=0.0173$; see
Fig.~\ref{fig:fig1}) when, for this metallicity, stars more massive
than $1.073 {{\cal M}_\odot}$ are predicted to have sustained core convection
during MS evolution (see Fig.~\ref{fig:fig2}).  A much more reasonable
hypothesis would be that the $F_{\rm over}$--Mass relationship plotted in
Fig.~\ref{fig:fig1} can be used at all metallicities, provided that its
zero-point is adjusted, as appropriate, to take into account the $Z$-dependence
of the transition mass.  Hence, a shift of the locus plotted in 
Fig.~\ref{fig:fig1} horizontally to the left by $0.064 {{\cal M}_\odot}$
would, for instance, yield the relation between $F_{\rm over}$ and mass to be
used for stars having $Z=0.04$.  This is, in fact, the procedure that has been
used in this investigation to determine the value of $F_{\rm over}$ that applies
to MS stars of arbitrary mass and metal abundance.

Before confronting the models with other observational tests (see \S 4), it is
important to verify that isochrones based on this calibration of $F_{\rm over}$
are just as capable of reproducing the CMDs of M$\,$67, NGC$\,$6819, and
NGC$\,$7789 as those derived from evolutionary tracks in which the same value 
of $F_{\rm over}$ is assumed for all masses.  Recall that, in their study of 
M$\,$67, \citet{vs04} computed grids of evolutionary tracks for $F_{\rm over} =
0.0, 0.07, 0.12$, and 0.20, and discovered that the isochrones for $F_{\rm over}
= 0.07$ did the best job of matching the photometric data.  All of the tracks 
in each grid, for masses up to $1.5 {{\cal M}_\odot}$, assumed the same value
of the overshooting parameter.  Similarly, \citet{gvs98} found that the set of
models in which each track, for masses ranging up to $2.0 {{\cal M}_\odot}$,
was computed on the assumption of $F_{\rm over} = 0.55$ provided the best fit
to the NGC$\,$7789 CMD.  There is clearly an inconsistency in assuming very
different values of $F_{\rm over}$ for stars of the same mass in the two
studies, but the key point here is that M$\,$67 is much older than NGC$\,$7789.
Whereas, e.g., $1.25 {{\cal M}_\odot}$ stars are just about to leave the MS
in M$\,$67, they are well below the turnoff in NGC$\,$7789.  As a consequence,
the implications for the isochrones appropriate to NGC$\,$7789 of assuming a
value of $F_{\rm over}$ as high as 0.55 or as low as 0.07 for, say, $1.25
{{\cal M}_\odot}$ stars are expected to be quite minor.  This expectation is
confirmed in the following plots.

\begin{figure}
\plotone{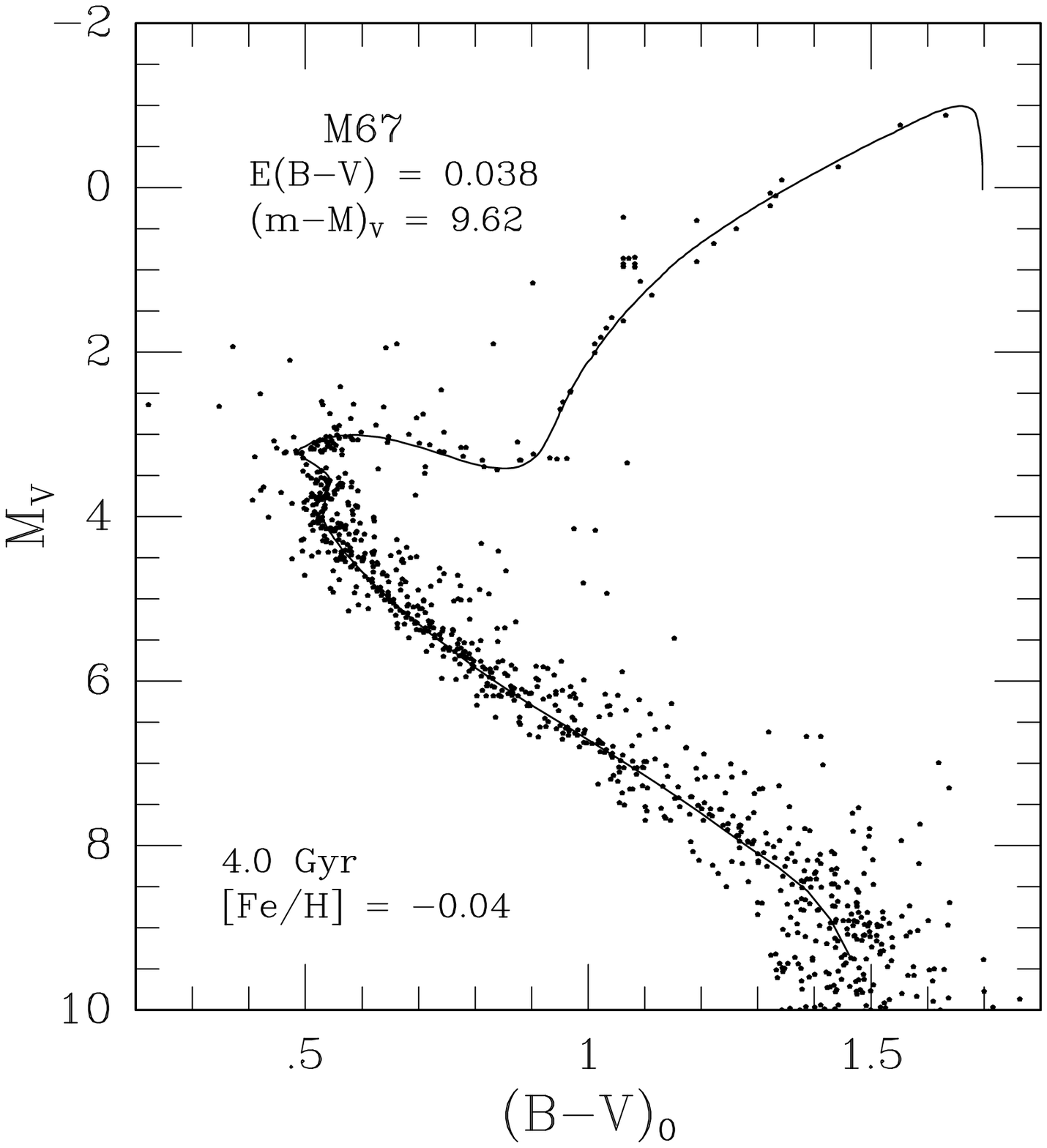}
\caption{Main-sequence fit of a 4.0 Gyr isochrone for [Fe/H] $=-0.04$ to the
 $BV$ photometry of M$\,$67 by Montgomery et al.~(1993), on the assumption of
 the indicated reddening and apparent distance modulus.} 
\label{fig:fig3}
\end{figure}

As shown in Figure~\ref{fig:fig3}, isochrones derived from evolutionary
tracks that employ the relation between $F_{\rm over}$ and mass plotted in
Fig.~\ref{fig:fig1} provide a superb match to the CMD of M$\,$67 given
by \citet{mmj93}.  Moreover, there are no obvious differences between the
overlay of the models onto the MS, TO, and subgiant stars shown here and that 
reported by VandenBerg \& Stetson (2004; see their Fig.~3e).  Of course, the
close agreement between theory and observations in this particular case is to
be expected given that the same M$\,$67 observations were used by \citet{vc03}
to constrain their color--$T_{\rm eff}$ relations (which have been adopted
throughout this investigation) for near solar abundance stars.\footnote{To be
specific, VC03 opted in favor of model-atmosphere-based color transformations
(e.g., \citealt{bg89}; \citealt{cas99}) for $\sim 5500$ K and hotter stars, but
they applied whatever corrections were necessary to purely synthetic colors in
order to achieve a good match of stellar models to both the MS and the red-giant
branch (RGB) of M$\,$67.  This procedure is readily justified by the fact that
the resultant $(B-V)$--$T_{\rm eff}$ relation agrees extremely well with the
latest empirical relation for field dwarfs having [Fe/H] $\approx 0.0$
(\citealt{sf00}).  Furthermore, because the predicted effective temperatures
for the cluster giants are well within the uncertainties of those inferred from
empirical $(V-K)$--$T_{\rm eff}$ relations (\citealt{bcp98}; \citealt{vb99}),
the $(B-V)$--$T_{\rm eff}$ relations applicable to low-gravity stars must be
close to those adopted by VC03 in order for the predicted $(V-K)$--$(B-V)$
diagram to be consistent with that observed.} 

\begin{figure}
\plotone{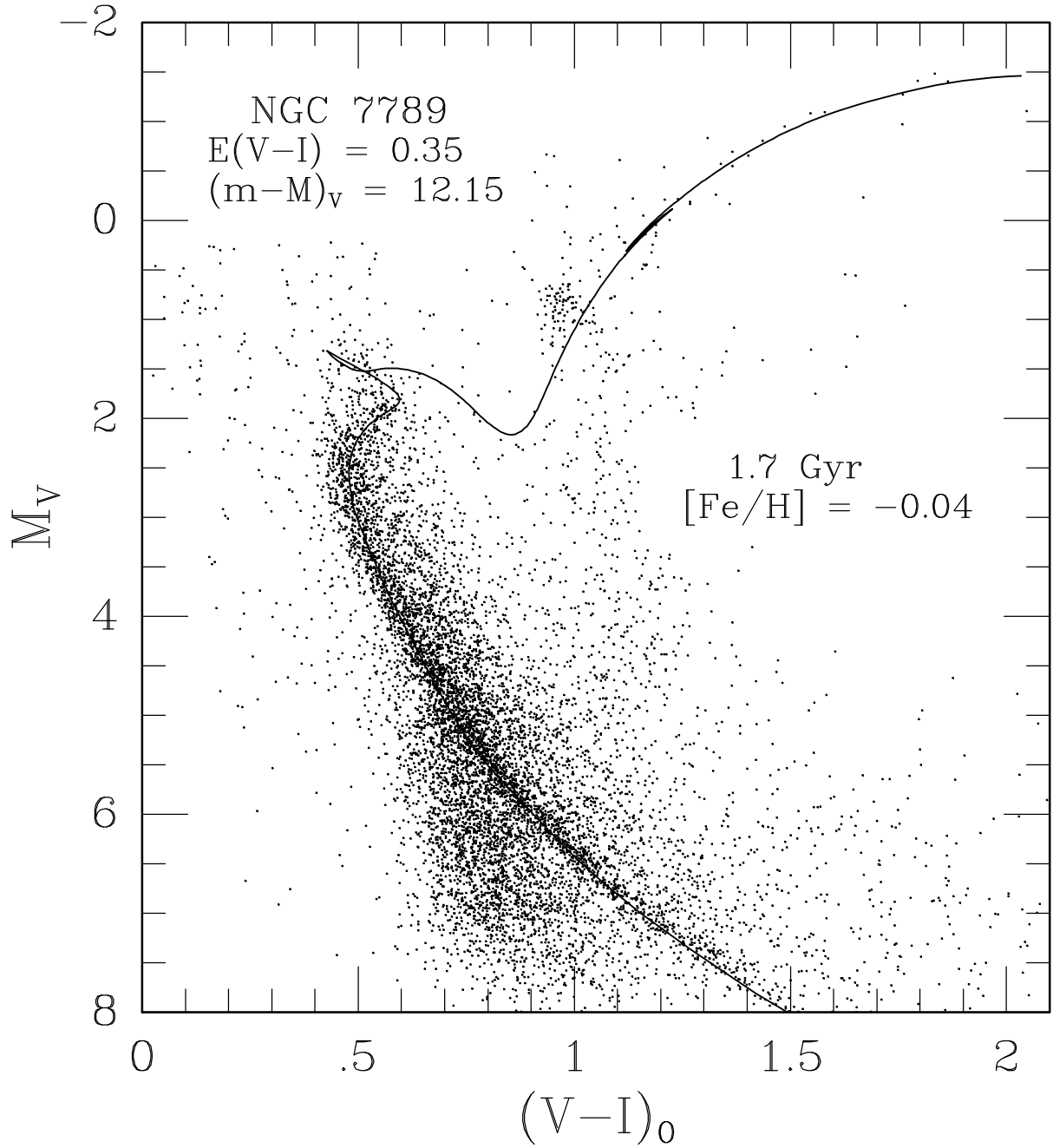}
\caption{Main-sequence fit of a 1.75 Gyr isochrone for [Fe/H] $=-0.04$ to the
 $VI$ photometry of NGC$\,$7789 by Gim et al.~(1998), on the assumption of the
 indicated reddening and apparent distance modulus.}
\label{fig:fig4}
\end{figure}

Figure~\ref{fig:fig4} illustrates that a 1.7 Gyr isochrone for [Fe/H]
$=-0.04$, which was interpolated from the same grid of evolutionary tracks used
to generate the 4.0 Gyr isochrone in the previous figure, provides a very good
match to the observations of NGC$\,$7789 obtained by \citet{gvs98}.  Because
the reddening to this cluster is quite uncertain, with estimates ranging from
$E(B-V) = 0.22$ to 0.32 (see Table 1 in the Gim et al.~paper), it is hard to
say whether the observations have been fitted to the right isochrone.  In our
limited exploration of parameter space, we have found that it is possible to
fit the turnoff data comparably well with either younger or older isochrones by
$\sim\pm 0.2$ Gyr (provided that somewhat different reddenings and distances
are assumed), though the best match to the {\it entire} MS, as well as to the
RGB, was found only if the cluster parameters are close to those indicated in
Fig.~\ref{fig:fig4}.  (Very similar fits to the cluster CMD were
reported by Gim et al. using isochrones for [Fe/H] $=0.0$. They did not
consider stellar models for [Fe/H] $=-0.04$ in their investigation.)

\begin{figure}
\plotone{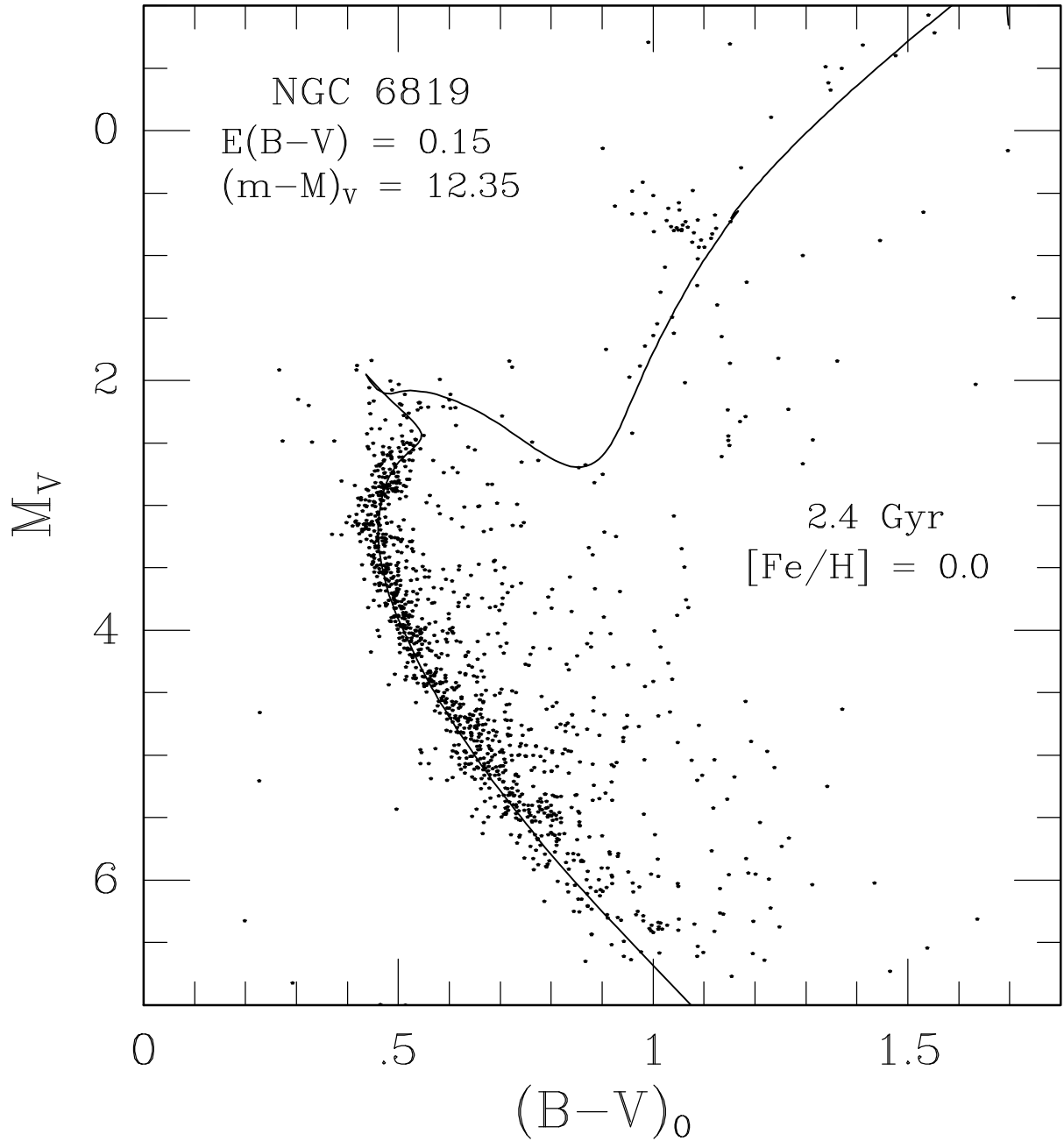}
\caption{Main-sequence fit of a 2.4 Gyr isochrone for [Fe/H] $=0.0$ to the
 $BV$ photometry of NGC$\,$6819 by Rosvick \& VandenBerg (1998), on the 
 assumption of the indicated reddening and apparent distance modulus.}
\label{fig:fig5}
\end{figure}

The reddening of NGC$\,$6819 seems to be quite well established at $E(B-V) =
0.14$--0.15 mag (see \citealt{rv98}; \citealt{bcg01}), implying an apparent
distance modulus of $\approx 12.35$ if derived from a main-sequence fit of
the photometry reported by Rosvick \& VandenBerg to our models for [Fe/H]
$=0.0$ (see Figure~\ref{fig:fig5}).  As discussed above, this cluster
appears to be slightly more metal rich than the Sun according to the latest
spectroscopic work.  However, we find that a 2.5 Gyr isochrone for [Fe/H]
$=-0.04$ actually provides the best match to the entire CMD --- if $E(B-V)
=0.15$ and $(m-M)_V = 12.30$.  It is evident from Fig.~\ref{fig:fig5}
that a solar-metallicity isochrone for 2.4 Gyr reproduces the cluster TO and MS
data quite well, though the theoretical giant branch is slightly to the red of
the observed one (possibly indicating a preference for a lower reddening).
The discrepancies are even larger if we assume that [Fe/H] $=+0.13$, which is
the next highest metallicity in our grids of stellar models.  In any case,
small differences in the adopted reddening or metallicity do not affect the
quality of the isochrone fits in the vicinity of the turnoff.  These indicate
that the models {\it are} allowing for approximately the right amount of CCO.

The main conclusion to be drawn from Figs.~3--5 is that our simple prescription
for $F_{\rm over}$ as a function of mass (see Fig.~1) works well for near
solar abundance stars.  Additional constraints on this calibration of CCO are
discussed in \S 4, where, in particular, data for a few low-metallicity systems
are examined to test our hypothesis concerning the probable dependence of the 
$F_{\rm over}$--Mass relation on metal abundance.

\section{The Stellar Evolutionary Models and Isochrones}
\label{sec:models}
The code described by \citet{vsr00} has been used to compute all of the stellar
evolutionary sequences that are presented in this investigation.  Even though
there have been some improvements to the basic physics of stars (notably, to the
conductive opacities; see \citealt{pbh99}), they have not yet been fully
implemented in the Victoria code.\footnote{Preliminary calculations of tracks
for $1.0 {{\cal M}_\odot}$ model stars having $Z=0.0001$ and $Z=0.01$ indicate
that the new conductive opacities by Potekhin et al.~(1999; also see
www.ioffe.ru/astro/conduct) lead to larger core masses at the tip of the giant
branch by only 0.005--$0.006 {{\cal M}_\odot}$ compared with the results
obtained using \citet{hl69} conductive opacities.   Such a small change in the
He core mass has only minor consequences for the luminosity of the horizontal
branch.}  In fact, since the main goal of the present paper is to extend the
grids of models reported by VandenBerg et al.~to sufficiently high masses
that isochrones can be calculated for any age in the range of 1 to 16 Gyr, it
is very important, for consistency reasons, that the same evolutionary program
be used.  Note that any models in the earlier work that had convective cores
which lasted throughout the MS phase were recomputed, taking CCO into account
according to the method described in \S 2.  All other low-mass tracks from the
previous study were appended to the new computations for higher masses,
resulting in grids of evolutionary sequences that, for each assumed chemical
composition, encompass a range in mass from 0.5 to $\sim 2.0$--$2.2 {{\cal
M}_\odot}$.

It should also be appreciated that diffusive processes have not been treated.
However, models that allow for gravitational settling and radiative
accelerations appear to be ruled out by recent observations of the
chemical abundances in TO and lower RGB stars in GCs (see \citealt{gra01}),
and by the lack of any variation of Li abundance with $T_{\rm eff}$ in warm
field halo dwarfs (see \citealt{rnb99}), {\it unless} some competing process,
such as turbulence, is invoked (\citealt{rmr02}; \citealt{vrm02}).
Helioseismic studies (\citealt{cpt93}; \citealt{trm98}) notwithstanding,
non-diffusive models are generally more consistent with both spectroscopic and
photometric data for old star clusters than those that allow for diffusion (for
reasons that are not currently understood).  Thus, there is considerable
justification to continue using the former in stellar population studies, and
to simply reduce the ages inferred from them by $\sim 10$\% (or less, if the
object under consideration has near solar abundances; see \citealt{mrr04}) if
the expected effects of diffusion on stellar ages, which are due mainly to the
settling of helium in the cores of stars, are taken into account.  (Otherwise,
for instance, color--$T_{\rm eff}$ relations would have to be fine-tuned in
order for diffusive isochrones to provide adequate fits to observed CMDs.)

The adopted chemical abundances for the model grids presented in this study are
listed in Table~\ref{tab:tab1}.  For each [Fe/H] value and mass-fraction
abundance of helium, $Y$, listed in the first and second columns, respectively,
$\alpha$-element abundances corresponding to [$\alpha$/Fe] $= 0.0, 0.3$, and
0.6 were assumed. [The $\alpha$ elements include O, Ne, Mg, Si, S, Ar, Ca, and
Ti.  The abundances of a few other elements (Na, Al, P, Cl, and Mn) were assumed
to be either correlated or anticorrelated with those of the $\alpha$ elements to
better represent their values in metal-poor stars.  Some justification for the
particular choices that have been made is provided by \citet{vsr00}, who also
tabulate (on the scale $\log\,{\rm N_H} = 12.0$) the adopted heavy-element
mixtures for each of the three cases considered.]  The mass-fraction abundances
of all elements heavier than helium, $Z$, are listed under the relevant
[$\alpha$/Fe] heading, along with the name of the file containing the tracks
in the form used by the interpolation software to generate isochrones,
luminosity functions, and color functions.  [Using an auxilliary code, the
model sequences originally computed were divided into so-called ``equivalent
evolutionary phase" points.  It is the output of that code; i.e., the ``.eep"
files that are used in the interpolation scheme and provided to interested
users along with the interpolation software --- see below and \citet{bv01}.]
For instance, ``vr0a-231.eep" contains the tracks for [Fe/H] $=-2.31$, assuming
zero enhancement in the $\alpha$-element abundances with respect to the solar
mix, while ``vr2a-231.eep" and ``vr4a-231.eep" contain the tracks for the same
[Fe/H], but, in turn, with two and four times the solar $\alpha$/Fe number
abundance ratio.

As \citet{vsr00} did not provide any models for [Fe/H] $> -0.30$, those listed
in Table~\ref{tab:tab1} for higher metallicities were newly computed for the
entire mass range considered.
For this study, we also decided to compute several sets of models
that could be used for stars and stellar populations having metal abundances in 
the range $-0.6\le$ [Fe/H] $\le +0.5$ and ages from $\approx 200$ Myr to 18 Gyr.
The adopted [Fe/H], $Y$, and $Z$ values, together with the names of the ``.eep"
files containing the tracks, are listed in Table~\ref{tab:tab2}.  For these
grids, the helium abundance varies with $Z$ according to $Y = 0.23544 + 2.2\,Z$,
and the heavy elements are assumed to have relative number fractions given by
the \citet{gn93} solar mix.  Both the helium enrichment law and the adopted
solar mix are slightly different for the [$\alpha$/Fe] $=0.0$ sets of models
whose properties are summarized in Table~\ref{tab:tab1} (see the VandenBerg et
al.~paper).  Due to the effects of the different heavy-element mixtures (mainly
on the opacity), the solar normalization for the sets of models listed in
Tables~\ref{tab:tab1} and~\ref{tab:tab2} differ slightly.  In the former case, 
the Standard Solar Model had $Y=0.2715$ and $\alpha_{\rm MLT} = 1.89$, where
$\alpha_{\rm MLT}$ is the usual mixing-length parameter, whereas the Standard
Solar Model in the latter case required $Y=0.2768$ and $\alpha_{\rm MLT}=1.90$.

Before turning to a brief summary of the refinements that have been made to the
interpolation software, it is worth pointing out that a forthcoming paper
(P.~D.~Dowler \& D.~A.~VandenBerg, in preparation) will review the main 
differences between overshooting and non-overshooting models, and discuss how
overshooting models that use the parameterized Roxburgh criterion to estimate
the amount of CCO in stars differ from those which simply extend the convective
core by some fraction of a pressure scale-height, $\Lambda$.  In particular, it
will show that the assumption of a constant value for $F_{\rm over}$ implies an
increasing value of $\Lambda$ over the course of a star's main-sequence 
evolution; and hence, that the two parameters are not equivalent.  (A more
extensive analysis, than that provided here, of the need for CCO in
intermediate-mass stars from observations of open clusters and binaries will
also be presented.)

\subsection{Isochrones and Luminosity/Color Functions}
\label{subsec:iso}
The software implemented to interpolate isochrones and, in conjunction with an
assumed initial mass function, to predict the numbers of stars along them, uses
the same general approach described by \citet{bv92} and refined by \citet{bv01},
but with a few significant modifications.  There are two fundamental
reasons for this. The first is that the extension of the grids to higher
masses introduces problems in identifying equivalent evolutionary phases
(EEPs) in such a way that they define monotonic Mass--Age relations throughout
each grid of tracks.  [Recall that the interpolation scheme relies
on the morphology of the temporal derivative of the effective temperature
$d(\log T_{\rm eff})/d(\log t)$.  The grids published in BV01 extended, at
most, to tracks for $1.25{\cal M}_\odot$, so no extreme variations of
derivative morphology were encountered.]  The second is that, in some of the
grids, even when the primary EEPs define a monotonic Mass--Age relation, Akima
spline interpolation fails along the secondary EEPs that span the region just
past the blue hook, shortly after it develops in the tracks.  (Stars which
maintain convective cores throughout the MS phase evolve rapidly to the blue
when their core hydrogen is exhausted; and it is only after H shell burning is
established, as a result of the concomitant core contraction, that their tracks
reverse direction and the stars eventually become red giants.  This blue hook
morphology is not found in the tracks of stars that have radiative cores on
the main sequence.)

\begin{figure}
\plotone{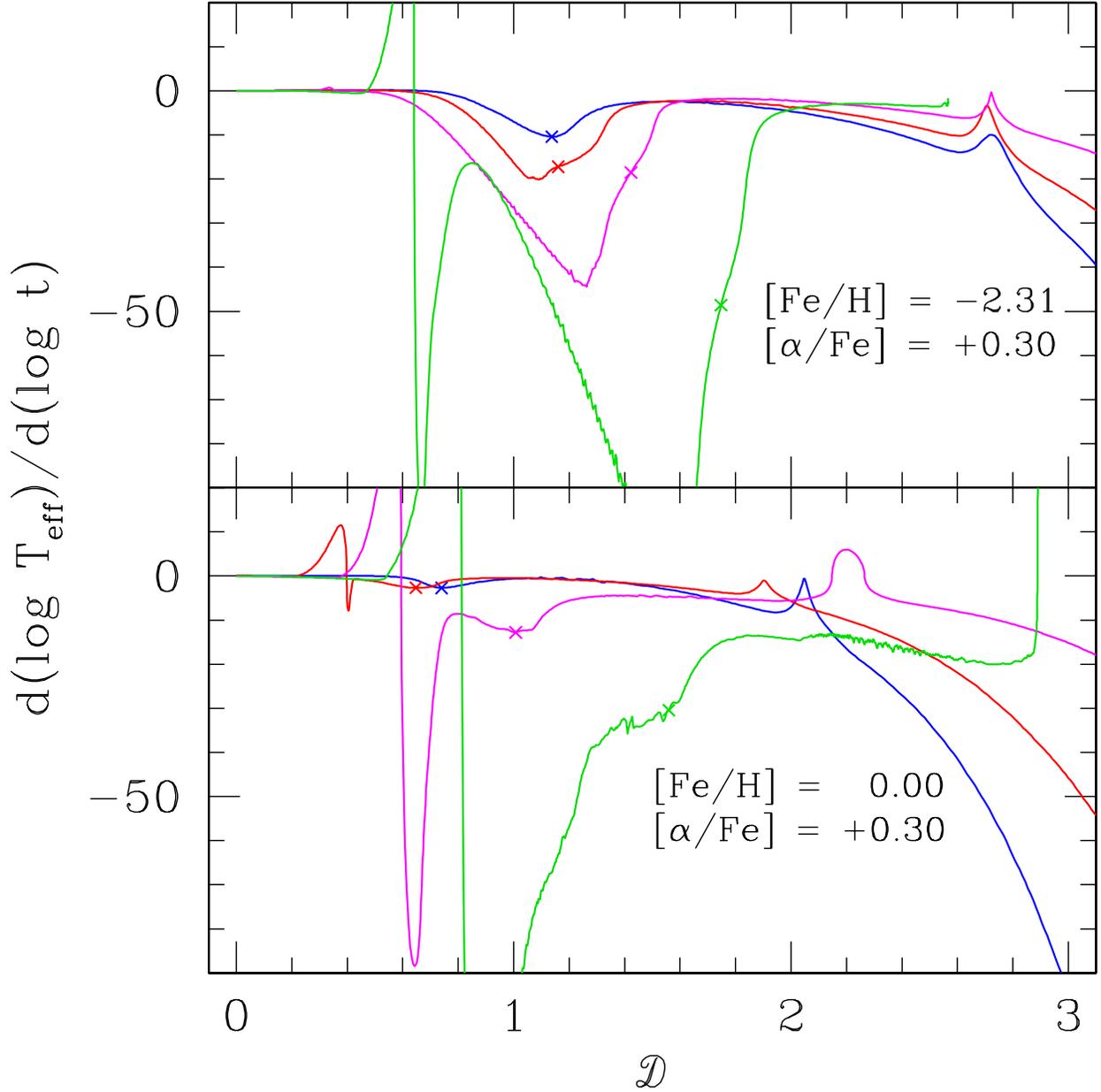}
\caption{The derivative d$(\log T_{\rm eff})$/d$(\log t)$ is plotted as a 
function of distance along an evolutionary track for the abundances shown.  
The distance is calculated from ${\cal D} = \sum\Delta{d}$ where $\Delta{d} =
[1.25\,(\Delta\log L)^2 +10.0\,(\Delta\log T_{\rm eff})^2]^{1/2}$ with
${\cal D} = 0.0$ at the lowest mass point on the isochrone; see BV01.)  In the
upper panel, the derivatives plotted correspond to those for 0.9 (blue),
1.1 (red), 1.5 (magenta) and $1.8{\cal M}_\odot$ (green) tracks; in the lower
panel, the derivatives correspond to 0.9 (blue), 1.4 (red), 1.8
(magenta), and $2.4{\cal M}_\odot$ (green) tracks. The adopted location
of the HZGP point is indicated by the symbol ${\bf\times}$ on each curve.}
\label{fig:fig6}
\end{figure}

Consider first the morphology of the temperature derivative in the metal-poor
tracks illustrated in the upper panel of Figure~\ref{fig:fig6}.  On the
$0.9{\cal M}_\odot$ track, the Hertzsprung gap point (HZGP) is easily
identifiable at the local derivative minimum.  However, on the $1.1{\cal
M}_\odot$ track, the development of apparently anomalous morphology at the
HZGP derivative minimum occurs even before the incipient blue hook becomes
evident in the $1.5{\cal M}_\odot$ track, on which the morphology of the
minimum is obviously different from that of the $0.9{\cal M}_\odot$ track. The
fully developed blue hook is clearly evident in the $1.8{\cal M}_\odot$ track,
which also shows a more extreme version of the transition to the giant branch.
These differences imply that the the minima do not correspond to equivalent
evolutionary phases: in the low-mass regime, the minimum is an artifact of the
rather diffuse pp-chain processing in the hydrogen-burning core, whereas in
the high-mass regime, it is an artifact of the more centrally concentrated
CNO cycle processes (due to their much higher temperature sensitivity) that
dominate the energy production.  The primary EEPs at the Hertzsprung gap and
near the base of the RGB (BRGB) bracket the bottom of the giant branch on
the $0.9{\cal M}_\odot$ track; the equivalent HZGP point on the higher mass
tracks is found at the inflection point in the derivative on the following side
of the minimum.

The lower panel of Figure~\ref{fig:fig6} illustrates that, in metal-rich
grids, the transition occurs in a distinctly different way.  In this instance,
the well-defined derivative minimum in the $1.8{\cal M}_\odot$ track, does
correspond to the HZGP EEP located at the local minimum in the $0.9{\cal
M}_\odot$ track; the corresponding primary EEP in the $2.4{\cal M}_\odot$
track is the inflection point marked on the following edge.

One other adjustment that had to be made to the interpolation scheme was to
manage the transition from stubby giant branches in the tracks of the more
massive stars, which terminate near the location of the evolutionary pause
(GBPS) that occurs when the H-burning shell contacts the chemical composition
discontinuity caused by the first dredge-up, to fully developed RGBs in the
lower mass tracks.  This was accomplished by making the GBPS primary EEP
degenerate with the RGB tip EEP for those tracks with stubby RGBs, such as the
$1.8{\cal M}_\odot$ track illustrated in the upper panel of
Figure~\ref{fig:fig6}, and in the $2.4{\cal M}_\odot$ track in the lower panel.

\begin{figure}
\plotone{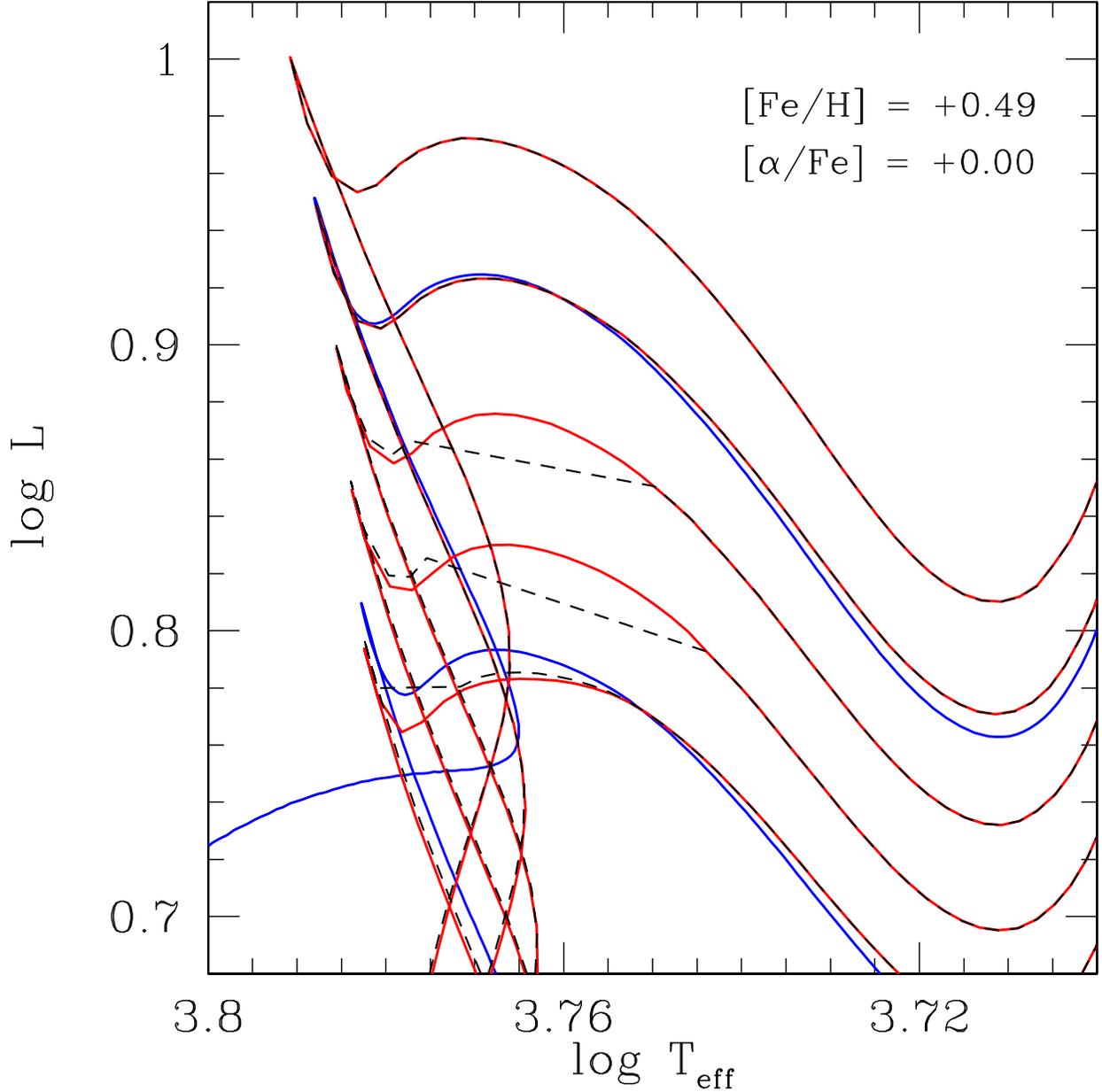}
\caption{The central H exhaustion and subgiant segments of the evolutionary
sequences for 1.4 and $1.5{\cal M}_\odot$ stars with the indicated abundances
are plotted (solid blue lines) along with several isochrones. The dashed black
lines represent the isochrones interpolated via the Akima spline for
the ages 2.6, 2.8, 3.0, 3.2, and 3.4 Gyr; the solid red lines represent the
same isochrones obtained via linear interpolation. The straight line segments
just past the blue ends of the hook in the 3.0, 3.2, and 3.4 Gyr isochrones
that were obtained using the Akima spline (the dashed black lines) indicate
the existence of gaps in the point distributions.  They correspond to mass
differences of slightly more than $0.001{\cal M}_\odot$.}
\label{fig:fig7}
\end{figure}

In \citet{bv92}, it was clearly demonstrated that linear interpolation works
extremely well with the prescribed set of primary EEPs defined therein in the
regime of low-mass stars. In BV01, Akima spline interpolation was introduced
primarily to improve the calculation of isochrone probability functions
(IPFs), luminosity functions, and color functions. However, in preparing the
current set of grids, we noticed that isochrones in the age range of
$\approx 2.8$--3.4 Gyr, derived via spline interpolation, occasionally
exhibited gaps in the point distribution over the transition from the blue
hook to the base of the RGB. As illustrated in Figure~\ref{fig:fig7}, the
same isochrones derived via linear interpolation do not have these gaps.  Our
interpretation of this is that, even though the EEP Mass-Age relations defined
by the primary and secondary EEPs at their locations on the evolutionary
tracks remain monotonic, very small wobbles of the spline (corresponding to
slightly more than $0.001{{\cal M}_\odot}$, at most) create ``forbidden" regions
in the interpolation relations {\it between} the tracks (i.e., zones where the
age increases slightly with increasing mass).  (Isochrone points interpolated
by the two methods occupy slightly different locations only because of
differences in the shape of the Mass--Age relations in the regions {\it between}
the tracks.  In both cases, the Akima spline was used to interpolate $\log L$
and $\log T_{\rm eff}$ between the tracks.)

\begin{figure}
\plotone{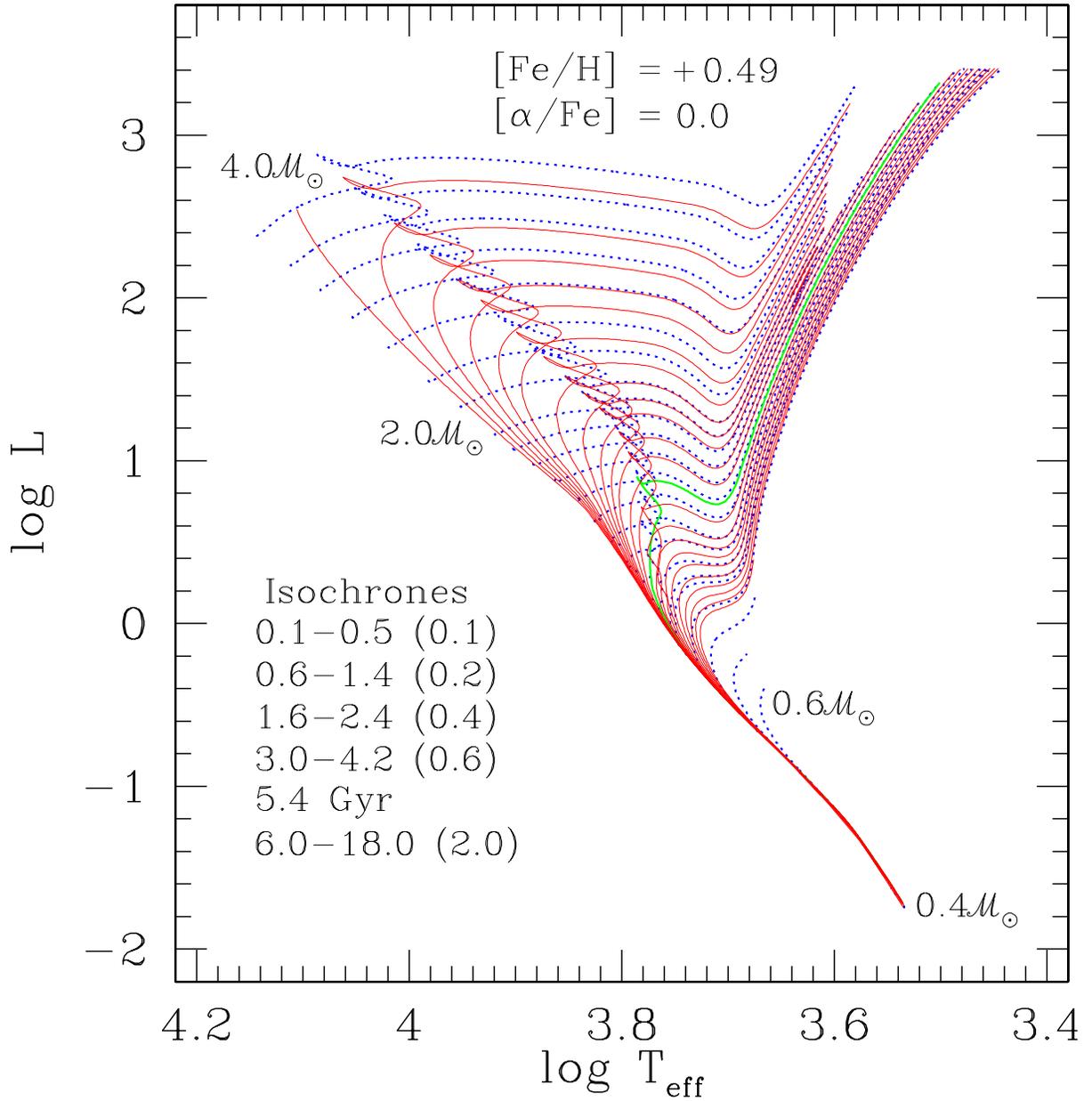}
\caption{The complete grid of evolutionary tracks for the scaled solar,
[Fe/H]$=+0.49$ grid are plotted with the dotted blue lines. The tracks
are plotted at $0.1{\cal M}_{\odot}$ intervals from 0.4 up to $2.0{\cal
M}_{\odot}$, then at $0.2{\cal M}_{\odot}$ intervals up to $2.4{\cal
M}_{\odot}$, and at $0.3{\cal M}_{\odot}$ intervals up to $3.6{\cal
M}_{\odot}$. There is an additional track at $1.031{\cal M}_{\odot}$, which is
the transitional mass at which the blue hook appears. The grid is
completed with a $4.0{\cal M}_{\odot}$ track.  The isochrones span the
range of ages from 0.1 to 18.0 Gyr (as indicated). Those plotted with
solid red lines were derived via spline interpolation; the single green
isochrone, with an age of 3 Gyr, was derived via linear interpolation.}
\label{fig:fig8}
\end{figure}

The complete grid of evolutionary sequences from which the segments of the 1.4
and $1.5 {{\cal M}_\odot}$ tracks plotted in the figure just discussed were
taken is shown in Figure~\ref{fig:fig8}.  Note that the spacing in mass is
quite fine, as a total of 25 tracks span the range in mass from 0.4 to $4.0
{{\cal M}_\odot}$.  This includes one track for $1.031 {{\cal M}_\odot}$, which
is the mass (for this particular grid) at which a transition is made between
stars that have radiative cores at central H exhaustion to those that maintain
convective cores throughout the MS phase.  (As a consequence of this transition,
higher mass tracks have blueward hooks at their turnoffs.)  Also plotted are
several isochrones to illustrate both the range in age for which isochrones
may be computed (0.2--18 Gyr) and the approximate region in the H--R diagram
where they are obtained using linear interpolation to avoid the problem
described in the preceding paragraph (i.e., in the vicinity of the 3.0 Gyr
isochrone, which has been plotted in green).  The main signatures of CCO are
(i) a widening of the main-sequence band (the region between the ZAMS and the 
line connecting the red ends of the hook feature on each of the higher mass
tracks), and (ii) a decrease in the mass at which a transition is made from an
extended to a stubby RGB.  In this particular grid, it occurs at $\approx 1.8
{{\cal M}_\odot}$.

\begin{figure}
\plotone{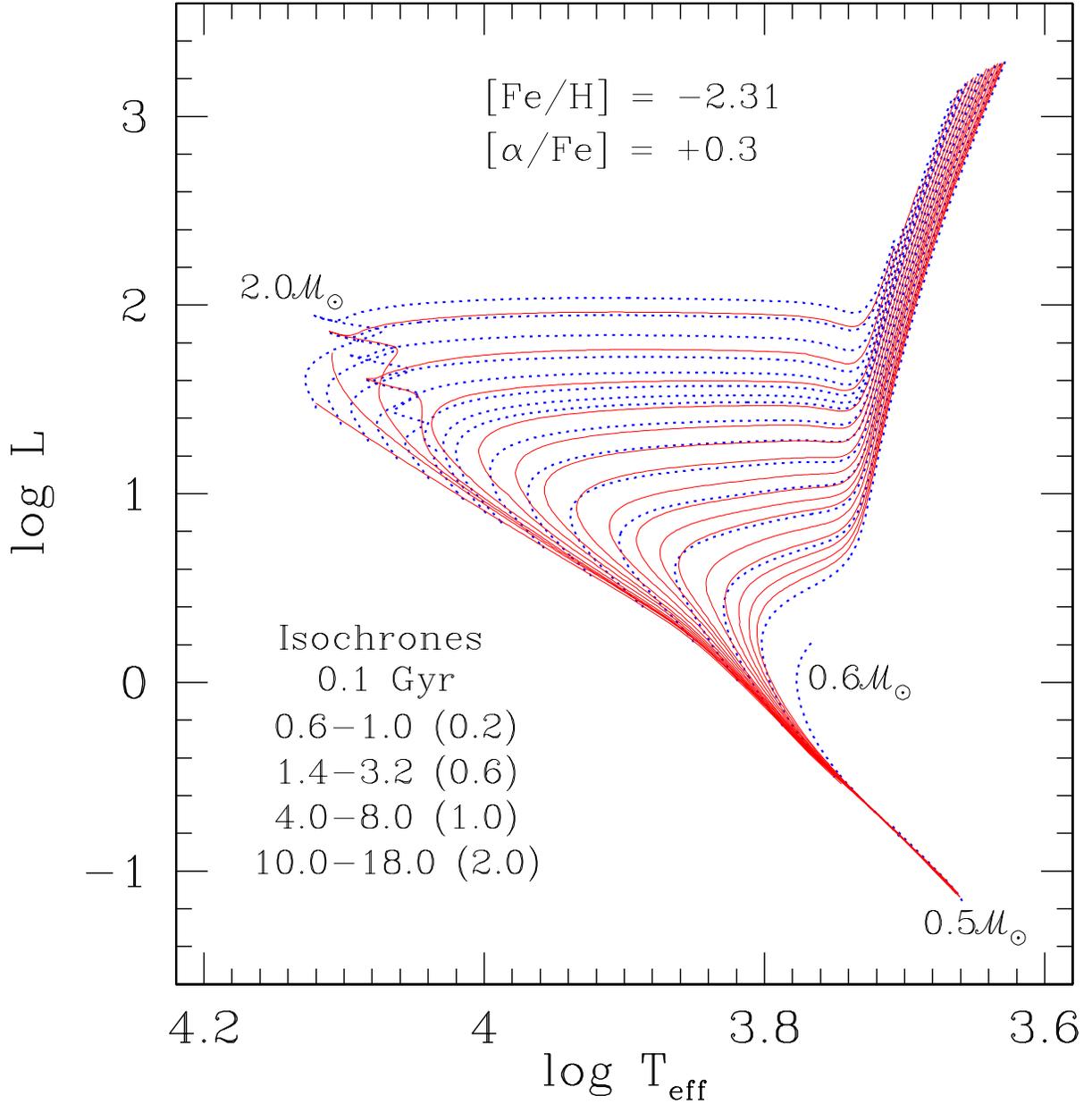}
\caption{The complete grid of evolutionary tracks for the
[Fe/H]$=-2.31$, $[\alpha/{\rm Fe}]=+0.3$ grid are plotted with the dotted
blue lines. The tracks are spaced at $0.1{\cal M}_{\odot}$ intervals
with an additional track at ${1.44\cal M}_{\odot}$, which is the
transitional mass for the development of the blue hook. The isochrones,
plotted as solid red lines, span the range of ages from 0.1 to 18.0 Gyr,
as indicated.}
\label{fig:fig9}
\end{figure}

All of the ``VRSS" grids (see Table~\ref{tab:tab2}) encompass the same ranges
in mass and age as that shown in Fig.~\ref{fig:fig8}.  In the case of the
grids that have been computed for [$\alpha$/Fe] $= 0.0, 0.3,$ and 0.6 (see
Table~\ref{tab:tab1}), tracks have been computed from $0.5 {{\cal M}_\odot}$
to that mass (within the range $2.0\le {{\cal M}_\odot} \le 2.2$) which enables 
isochrones to be computed down to $\sim 1$ Gyr.  An example of the latter is
given in Figure~\ref{fig:fig9}, which illustrates the evolutionary sequences
that comprise the ``VR2A-231" grid (i.e., that for [$\alpha$/Fe] $=0.3$ and
[Fe/H] $=-2.31$), along with representative isochrones.  Note that the
transition between stars with, and without, blueward hooks at their turnoffs
occurs at a much higher mass ($1.44 {{\cal M}_\odot}$) than in the grid plotted
in the previous figure.

\section{Observational Tests of the Stellar Models}
\label{sec:obs}

\subsection{Metal-Deficient Star Clusters}
\label{subsec:mpoor}
In principle, the many young, metal-poor star clusters that populate the
Large Magellanic Cloud (LMC) should provide especially tight constraints on the
extent of overshooting in low-metallicity stars.  However, in practice, this
is not yet possible, mainly because current estimates of their chemical
compositions are so uncertain.  For instance, even though very well-defined
CMDs are now available for NGC$\,$2155, NGC$\,$2173, and SL$\,$556 (see
\citealt{gal03}), it is not at all clear which models should be compared with
the photometric data when current [Fe/H] determinations for each of these
systems vary by about 0.6 dex (typically from [Fe/H] $\sim -1.2$ to $\sim 
-0.6$; see the review of the properties of these clusters by \citealt{bng03}).
All that one can reasonably do, until this situation improves, is ascertain
whether it is possible to find an isochrone (or generate a synthetic CMD) that
does a satisfactory job of matching an observed CMD on the assumption of a
reddening, distance, and metal abundance within the ranges of their
uncertainties.  This is effectively what was done by Bertelli et al.~and by
\citet{wgd03} in their studies of NGC$\,$2155, NGC$\,$2173, and SL$\,$556.

\begin{figure}
\plotone{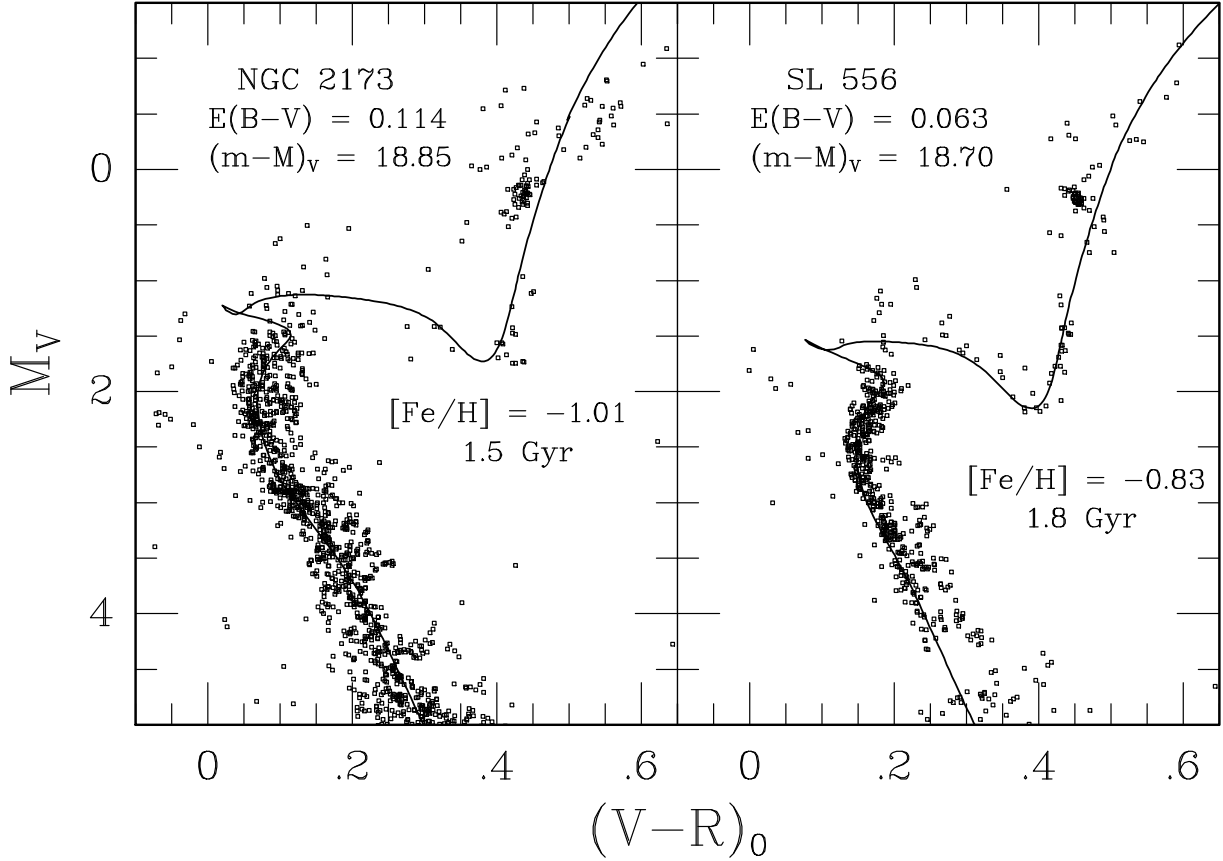}
\caption{Overlay of those isochrones for the indicated [Fe/H] values and ages
that best reproduce the CMDs of the LMC clusters NGC$\,$2173 and SL$\,$556
(Gallart et al.~2003) on the assumption of the reddening and apparent distance
modulus given in the upper left-hand corner of each panel (see the text).  The
abundances of the $\alpha$-elements are assumed to be enhanced by [$\alpha$/Fe]
$=0.3$.}
\label{fig:fig10}
\end{figure}

\begin{figure}
\plotone{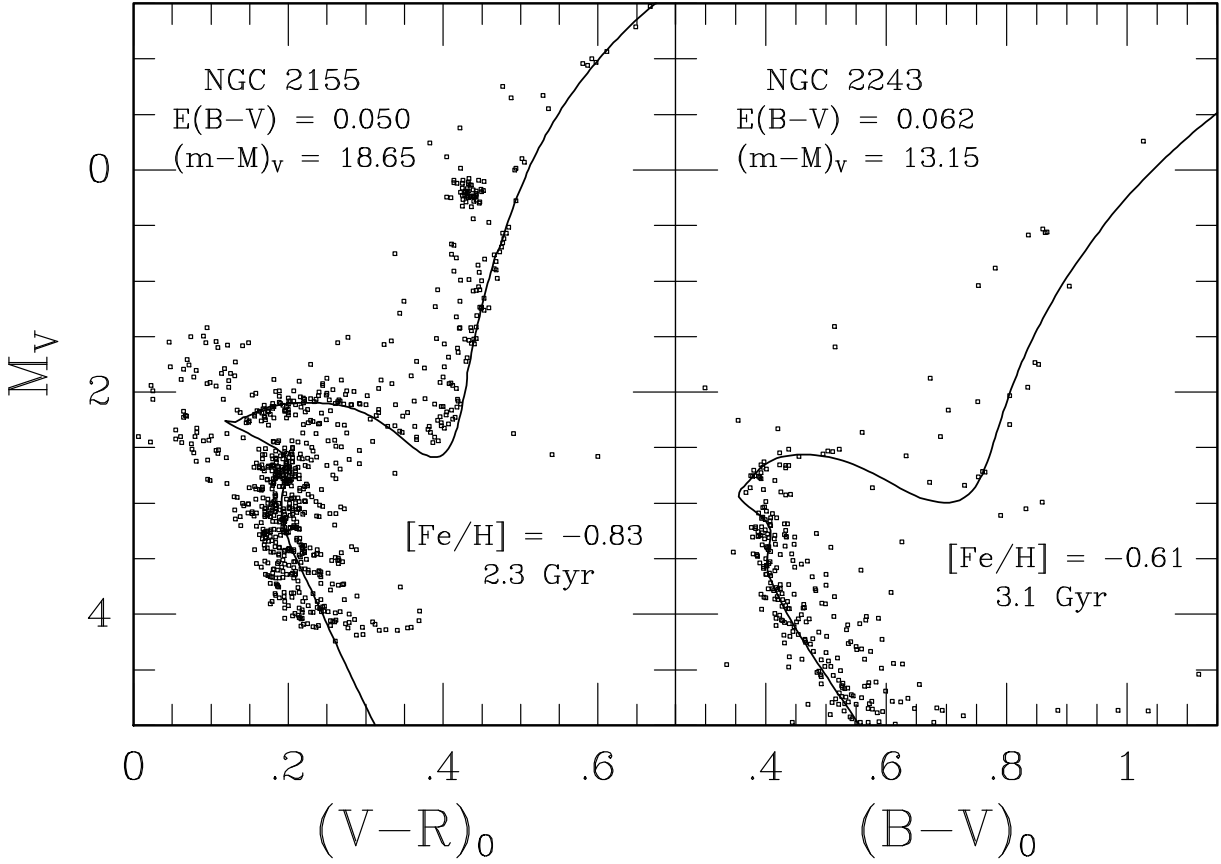}
\caption{Similar to the previous figure, except that isochrones are compared
with Gallart et al.~(2003) $VR$ observations of NGC$\,$2155 (left-hand panel),
also in the LMC, and with $BV$ photometry for the Galactic open cluster
NGC$\,$2243 (right-hand panel) by Bergbusch et al.~(1991).  In the latter case,
the adopted distance modulus was derived from a main-sequence fit of the
observed CMD to the models on the assumption of the indicated reddening (see
the text).  In both cases, enhanced abundances of the $\alpha$-elements by
[$\alpha$/Fe] $=0.3$ has been assumed.}
\label{fig:fig11}
\end{figure}

How well are our isochrones able to fit the CMDs of these three clusters?  It
is certainly of some interest to know the answer to this question --- and, as
shown below, our models appear to fare quite well.  Our approach is to adopt
the line-of-sight reddenings from the \citet{sfd98} dust maps and to assume,
as a first approximation, that all three systems are at a distance corresponding
to $(m-M)_0 =18.50$, which is very close to the mean LMC modulus that has been
derived using many different techniques (e.g., see \citealt{ben02}).  Having
set the reddening and distance, there are no free parameters other than the
metallicity (and age).  The latter can be ``determined" simply by overlaying 
the isochrones for different chemical compositions and ages onto an observed
CMD and then selecting that isochrone which most closely matches the
observations.  (The metal abundance obtained in this way is obviously nothing
more than a ``prediction" and, as such, of limited usefulness only when, as
in the case of the LMC clusters, metallicities are poorly known.)  The result
of this exercise is shown in Figure~\ref{fig:fig10} and in the left-hand
panel of Figure~\ref{fig:fig11}.  [Note that $E(V-R) = 0.59\,E(B-V)$ has been
assumed; see, e.g., Bessell et al.~(1998).  Furthermore, [$\alpha$/Fe] $=0.3$
has been adopted simply because such an enhancement has been found in the
majority of stars in the Milky Way with [Fe/H] $\lta -0.6$, whether in globular
clusters (e.g., \citealt{car96}; \citealt{kss98}) or in the field (e.g.,
\citealt{zm90}; \citealt{ns97}).  The accuracy of this assumption for the 
three LMC clusters considered has not yet been established.]

The level of agreement between theory and observations is actually surprisingly
good (and quite comparable to the findings of \citealt{bng03}; \citealt{wgd03}).
[The most obvious discrepancy is the failure of the isochrones to match the
$V-R$ colors of the upper RGB stars in NGC$\,$2173 and SL$\,$556.  This may
indicate that the inferred metallicities and/or distances should be revised
slightly or that there is a problem with the model temperatures and/or the
$(V-R)$--$T_{\rm eff}$ relations that we have used (from VC03) to transpose
the isochrones to the observed plane.  In fact, there is already some evidence
to suggest that the $V-R$ colors given by the VC03 transformations are too
blue for low-gravity stars --- see the study of NGC$\,$188 by \citet{vs04}.
However, until the observational parameters are placed on a firmer footing,
it would be premature to conclude that the latter explanation is the correct
one.  NGC$\,$2155 would not show a similar mismatch between the predicted and
observed giant branch if, e.g., the adopted distance or metallicity was too
high.  The fact that the MS segment of the isochrone used to fit the
observations is too red does provide some indication that a lower metal
abundance may be more appropriate, but we have found that our models for
[Fe/H] $=-1.01$, which is the next lowest metallicity in our grids, are too
blue.  This suggests some preference for models with [Fe/H] $\sim -0.9$.]

The inferred [Fe/H] values are all well within the ranges of uncertainty
encompassing current photometric and spectroscopic metallicity estimates and,
in fact, are close to the metal abundances deduced or assumed by Bertelli et
al.~and Woo et al.  Even our derived ages are similar to those found by the
latter: what differences exist can be attributed mostly to the different
assumptions that have been made regarding the cluster reddenings and distances.
Indeed, until much tighter constraints are placed on the cluster metal
abundances, it is not possible to say whose interpretations of the observed
CMDs are the most accurate ones.  Certainly, from our perspective, it is very
encouraging that NGC$\,$2155, NGC$\,$2173, and SL$\,$556 do not appear to pose
any serious difficulties for the isochrones presented in this investigation,
which is really the main point of including a brief analysis of data for these
systems in this investigation.

A much better test of the overshooting models is provided by the metal-poor
Galactic cluster NGC$\,$2243.  Fortunately, \citet{aat05} have just carried
out a very thorough analysis of $uvby\,Ca\,H_\beta$ photometry for this 
system, from which they found $E(B-V) = 0.055$ and [Fe/H] $=-0.57$.  Their
reddening estimate is consistent with that derived from the Schlegel et
al.~(1998) dust maps, which give $E(B-V) = 0.074$.  As Anthony-Twarog et
al.~point out, the latter provides a line-of-sight upper limit, though it
should be close to the value appropriate for NGC$\,$2243 given that this
cluster lies above the Galactic plane.  Moreover, from their examination
of the work that has been carried out on this cluster to date, they suggest
that both photometric and spectroscopic metallicity determinations seem to be
converging to an [Fe/H] value of $\approx -0.55\pm 0.1$.

As illustrated in the right-hand panel of Fig.~\ref{fig:fig11}, an
overshooting isochrone for 3.1 Gyr, [Fe/H] $=-0.61$, and [$\alpha$/Fe] $=0.3$
provides an excellent match to the CMD obtained by \citet{bvi91}.  This
metallicity differs from the Anthony-Twarog et al.~(2005) best estimate by
$< 0.1$ dex, and the assumed reddening is approximately midway between the
$E(B-V)$ value derived by Anthony-Twarog et al.~and that from the Schlegel et
al.~(1998) dust maps.  Insofar as the $\alpha$-element abundances are concerned,
\citet{gc94} obtained [$\alpha$/Fe] $\approx 0.1$ (if [Fe/H] $=-0.48$) from
high resolution spectroscopy of two cluster giants.  The adoption of an iron
abundance lower by $\sim 0.1$ dex would imply an increase in the [$\alpha$/Fe]
value by the same amount.  Importantly, Gratton \& Contarini have concluded
that a scaled-solar composition is not favored by their data.  Indeed, they
do not find any significant differences in the chemical compositions of
NGC$\,$2243 and field stars of the same [Fe/H].

One could, of course, get similar fits of isochrones to the observed CMD on
the assumption of slightly lower or higher reddenings, but the derived distance
moduli from the main-sequence fitting technique and the ages would also differ
slightly.  (Our impression is that the best match to the observations is 
obtained for the adopted parameter values.)  NGC$\,$2243 provides an especially
important check of our calibration of $F_{\rm over}$ because it is a
sufficiently old cluster that the upper MS stars have masses in the range
where the overshooting parameter is a strong function of mass.  The fact that
the models are able to reproduce the morphology of the cluster turnoff so well
provides good support for our $F_{\rm over}$--mass--metallicity relation.  
 
\subsection{The Binary TZ Fornacis}
\label{subsec:tzf}
TZ For is one of a very small group of binaries with sufficiently
well-determined masses, radii, effective temperatures, {\it and} metal
abundances that they provide stringent tests of stellar models.  Indeed,
it has proven to be especially difficult for computed models, even those
that allow for convective overshooting, to match the properties of TZ For
(see \citealt{acn91}; and, in particular, \citealt{lv02}).  According to
Andersen et al., the components of this binary have masses of $1.95 \pm 0.03$
and $2.05 \pm 0.06 {{\cal M}_\odot}$, radii of $3.96 \pm 0.09$ and $8.32 \pm
0.12 {{\cal R}_\odot}$, and effective temperatures corresponding to
$\log T_{\rm eff} = 3.803 \pm 0.007$ and $3.699\pm 0.009$, respectively.
They also obtained [Fe/H] $= -0.1\pm 0.1$ from a direct spectroscopic
determination of its metal abundance. 

\begin{figure}
\plotone{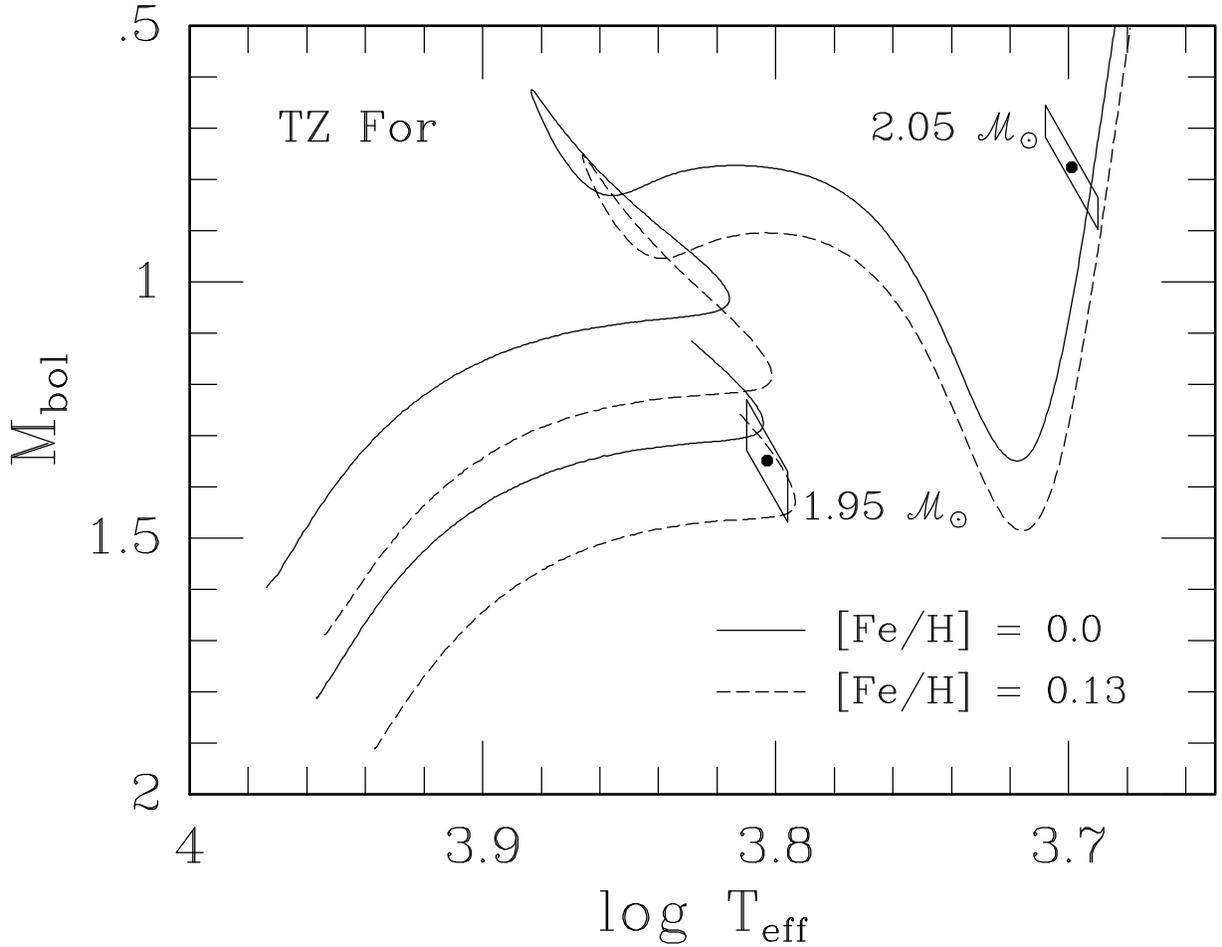}
\caption{Comparison of computed evolutionary tracks for the observed masses of
TZ Fornacis and for metallicities [Fe/H]$=0.0$ (solid curves) and
$+0.13$ (dashed curves) with the observed parameters of the binary
(filled circles and $1\sigma$ error parallelograms) obtained by
Anderson et al. (1998). From spectroscopic observations of TZ For, the
latter determined an [Fe/H]$=+0.1\pm 0.1$.  (A parallelogram is a more
accurate representation of the error boxes than a rectangle, as is
often used, because the uncertainty in $M_{\rm bol}$ at a given $\log
T_{\rm eff}$ is due only to the uncertainty in the stellar radius.)}
\label{fig:fig12}
\end{figure}

TZ For played no role in the calibration of the overshooting parameter,
$F_{\rm over}$, so it is particularly pleasing to find that our models are
able to provide quite a satisfactory fit to its components.  The locations of
these stars are plotted in Figure~\ref{fig:fig12} on the
($M_{\rm bol},\,\log T_{\rm eff}$)--plane, together with evolutionary tracks
that have been computed for the observed masses and two different
metallicities, [Fe/H] $=0.0$ and $+0.13$.  Taken at face value, the models
suggest that the fainter component has a metal abundance close to [Fe/H]
$=+0.06$, which is well within the uncertainty of the spectroscopically
determined metallicity.  If the actual iron abundance is somewhat higher or
lower than this value, one could still obtain a precise fit to the luminosity
and temperature of this star by adopting a helium abundance that differs
somewhat from the assumed values (see Table~\ref{tab:tab2}).

According to the models, the $1.95 {{\cal M}_\odot}$ star is nearing the end of
its main-sequence phase, with an age in the range of 1.26 Gyr (if [Fe/H] $=0.0$)
to 1.33 Gyr (if [Fe/H] $=+0.13$).  Its companion is predicted to have an age
at its observed luminosity between 1.12 and 1.19 Gyr (assuming [Fe/H] $=0.0$ and
$+0.13$, respectively) if it is a first-ascent red-giant-branch star.  Their
ages are not in perfect agreement, but the uncertainties in their masses, though
small, are still large enough that identical ages could almost certainly be
found for both stars if their masses are more nearly equal than observed
(and remain within the observed $1\sigma$ mass limits).  More likely is the
possibility that the most massive component is in the core He-burning phase
(see the discussion by \citealt{acn91}), in which case there may not be an age
problem.  (Unfortunately, we do not yet have the means to test this hypothesis,
but it goes without saying that the probability of finding a binary like TZ For
is much higher if both of its components are in long-lived evolutionary phases.)
In any case, we conclude from our analysis of this binary that an amount of
overshooting equivalent to $F_{\rm over} \approx 0.55$ {\it is} required to
model $\sim 2 {{\cal M}_\odot}$ stars having metallicities close to that of
the Sun.

\subsection{Globular Cluster Giant Branches}
\label{subsec:rgb}
\begin{figure}
\plotone{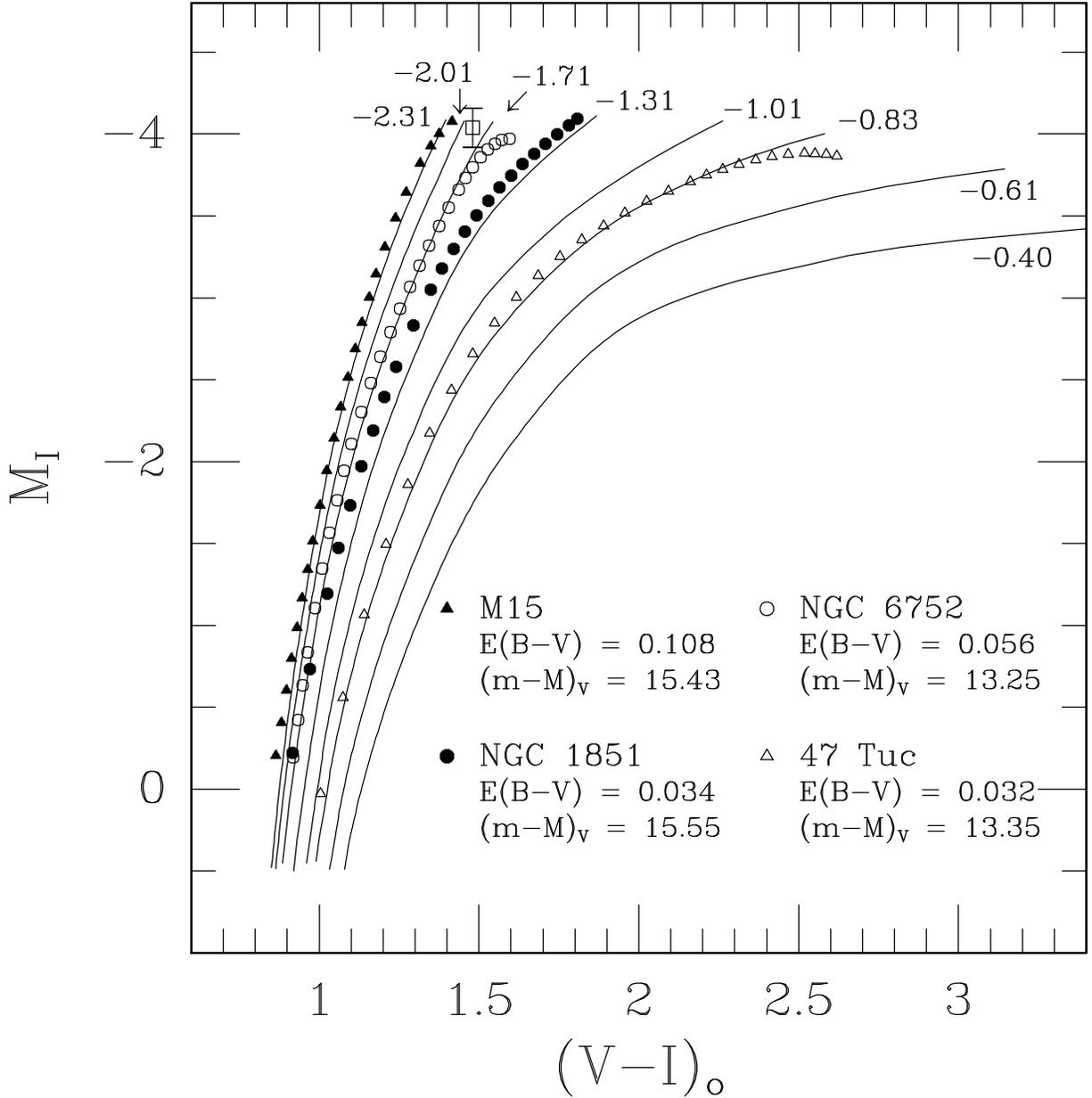}
\caption{Overlay of the fiducial sequences derived by Da Costa \& Armandroff
(1990) for M$\,$15, NGC$\,$6752, NGC$\,$1851, and 47 Tucanae, on the assumption
of the indicated reddenings and apparent distance moduli, onto the giant-branch
segments of 13 Gyr isochrones for [Fe/H] values ranging from $-2.31$ to $-0.40$,
assuming [$\alpha$/Fe] $=0.3$ in each case.  Note that, as shown by VandenBerg
(2000), turnoff ages close to 13 Gyr are consistent with the adopted $(m-M)_V$
values only for M$\,$15 and NGC$\,$6752: somewhat younger ages ($\approx 11.5$
Gyr) are obtained for NGC$\,$1851 and 47 Tuc.  The large open square, with error
bars, gives the absolute $I$ magnitude of the RGB tip stars in $\omega$ Centauri
determined by Bellazzini et al.~(2001); see the text.}
\label{fig:fig13}
\end{figure}

Vandenberg et al.~(2000) have already shown that their computed giant branches
for $\sim 0.8 {{\cal M}_\odot}$ metal-poor stars are in good agreement with
those derived for Galactic globular clusters (GCs) on the $(\log T_{\rm
eff},\,M_{\rm bol})$--plane by \citet{fpc81} using infrared photometry.
Moreover, BV01 have found that there is good consistency between the
models and the $V-I,\,I$ RGB fiducials for several GCs that were determined by
\citet{da90}, if Schlegel et al.~(1998) reddenings and distance moduli based on
zero-age horizontal-branch (ZAHB) models are assumed.  A very similar plot to
that provided in the latter study is given in Figure~\ref{fig:fig13},
except that a wider range in [Fe/H] is considered and the recent determination
by \citet{bfp01} of the absolute $I$ magnitude of the RGB tip ($M_I^{\rm TRGB}$)
of $\omega$ Centauri is shown (the {\it open square}).  Their estimate of
$M_I^{\rm TRGB} = 4.04\pm 0.12$, which is based on extensive cluster photometry
and the distance to $\omega$ Cen obtained by \citet{tkp01} from a detached
eclipsing binary, provides the most accurate {\it empirical} zero-point for the
$M_I^{\rm TRGB}$--[Fe/H] relationship that is currently available.

For the sake of clarity, the {\it open square} has been located at $(V-I)_0 =
1.48$, which is very close to the color of RGB tip stars having [Fe/H] $=-1.71$,
according to the relation between $(V-I)_0^{\rm TRGB}$ and [Fe/H] determined
by Bellazzini et al.~(2001).  (This will depend on the particular metallicity
scale which is assumed.)  Importantly, the absolute $I^{\rm TRGB}$ magnitudes
predicted by the metal-poor isochrones plotted in Fig.~\ref{fig:fig13}
(indeed, by those for all [Fe/H] values less than $\sim -0.8$) agree very well
with the empirical derivation of $M_I^{\rm TRGB}$.  Comparable agreement is
also found for the absolute $I^{\rm TRGB}$ magnitudes of M$\,$15, NGC$\,$6752,
and NGC$\,$1851 when distances based on VandenBerg et al.~(2000) ZAHB loci are
assumed (see VandenBerg 2000).  This consistency adds to the support for these
particular HB models found from recent studies of the pulsational properties
of RR Lyraes in GCs (see \citealt{dc99}; \citealt{ccc05}).  In fact, the
ZAHB-based distance modulus for NGC$\,$6752 is nearly identical with that
derived from its white dwarfs by \citet{rb96}; specifically, $(m-M)_V =
13.22\pm 0.1$ if $E(B-V) = 0.056$ and $A_V = 3.1\,E(B-V)$.  

The absolute $I^{\rm TRGB}$ magnitude of 47 Tuc is slightly fainter than
those of the other GCs considered in Fig.~\ref{fig:fig13}, but this is
implied by several of the latest distance determinations.  In particular,
\citet{zr01} obtained $(m-M)_V = 13.27\pm 0.14$ from the cluster white dwarfs,
while the fit of the main-sequence of 47 Tuc to that defined by local Population
II subdwarfs by \citet{gsa02} yielded $(m-M)_V = 13.33\pm 0.04\pm 0.1$.  (Note
that our estimate of the distance modulus of 47 Tuc assumes that the cluster has
[Fe/H] $=-0.83$: if this were increased to $-0.71$, the ZAHB-based modulus would
decrease to 13.30.)  One possible implication of these results is that the
variation of $M_I^{\rm TRGB}$ with [Fe/H] at higher metal abundances may be
slightly steeper than that found by Bellazzini et al.~(2001; their equation 4).

Perhaps the main concern with Fig.~\ref{fig:fig13} is that the cluster
metallicities, as inferred from the isochrones, tend to be on the low side of
most estimates (see Table~\ref{tab:tab3}).  It is not clear that this is
necessarily a problem with the models, but it is certainly possible that either
the predicted temperatures are too cool or the predicted colors are too red.
Although, as noted above, the model $T_{\rm eff}$ scale is consistent with
that derived for GCs from $V-K$ photometry, the uncertainties in the empirically
determined temperatures are $\sim\pm 100$ K.  This is easily more than enough
to reconcile the models with the latest spectroscopic abundances for several
of the clusters.

For instance, very close to the [Fe/H] values determined by \citet{ki03} for
NGC$\,$6752, NGC$\,$1851, and 47 Tuc would be implied by the superposition of
the cluster fiducials onto the isochrones if the temperatures of the latter were
increased by only 70 K.  (Somewhat larger adjustments would be needed if the
most accurate metallicities are those by \citealt{cg97}).  The difficulty with
this solution is that the implied metallicity of M$\,$15 would be $\approx
-2.1$, in conflict with their spectroscopic value of $-2.42$ (see
Table~\ref{tab:tab3}).  Without any justification for doing so, it would not
be very satisfying to correct the temperatures of the isochrones for some metal
abundances, but not for others.

A more optimistic comment on the comparison between theory and observations
presented in Fig.~\ref{fig:fig13} is that the predicted [Fe/H] scale is
consistent with that derived by \citet{zw84} to within $\approx 0.15$ dex.
This is comparable to, if not smaller than, the uncertainties associated with
most [Fe/H] determinations at the present time (e.g., compare the results
listed in Table~\ref{tab:tab3}).  Needless to say, it is going to be very
difficult to say whether or not the colors derived from the VC03 color--$T_{\rm
eff}$ relations are too red until much tighter constraints are placed on [Fe/H]
and temperature measurements.  Fortunately, model colors do not play a critical
role in the interpretation of the CMDs for simple stellar populations, like
open and globular star clusters.  It is mainly to address the observations of
those systems that have complicated star formation and chemical evolution
histories (e.g., the LMC, dwarf spheroidal galaxies) that some calibration of
the model colors is necessary (see the fairly extensive discussion of this
matter by \citealt{van05}).

\begin{figure}
\plotone{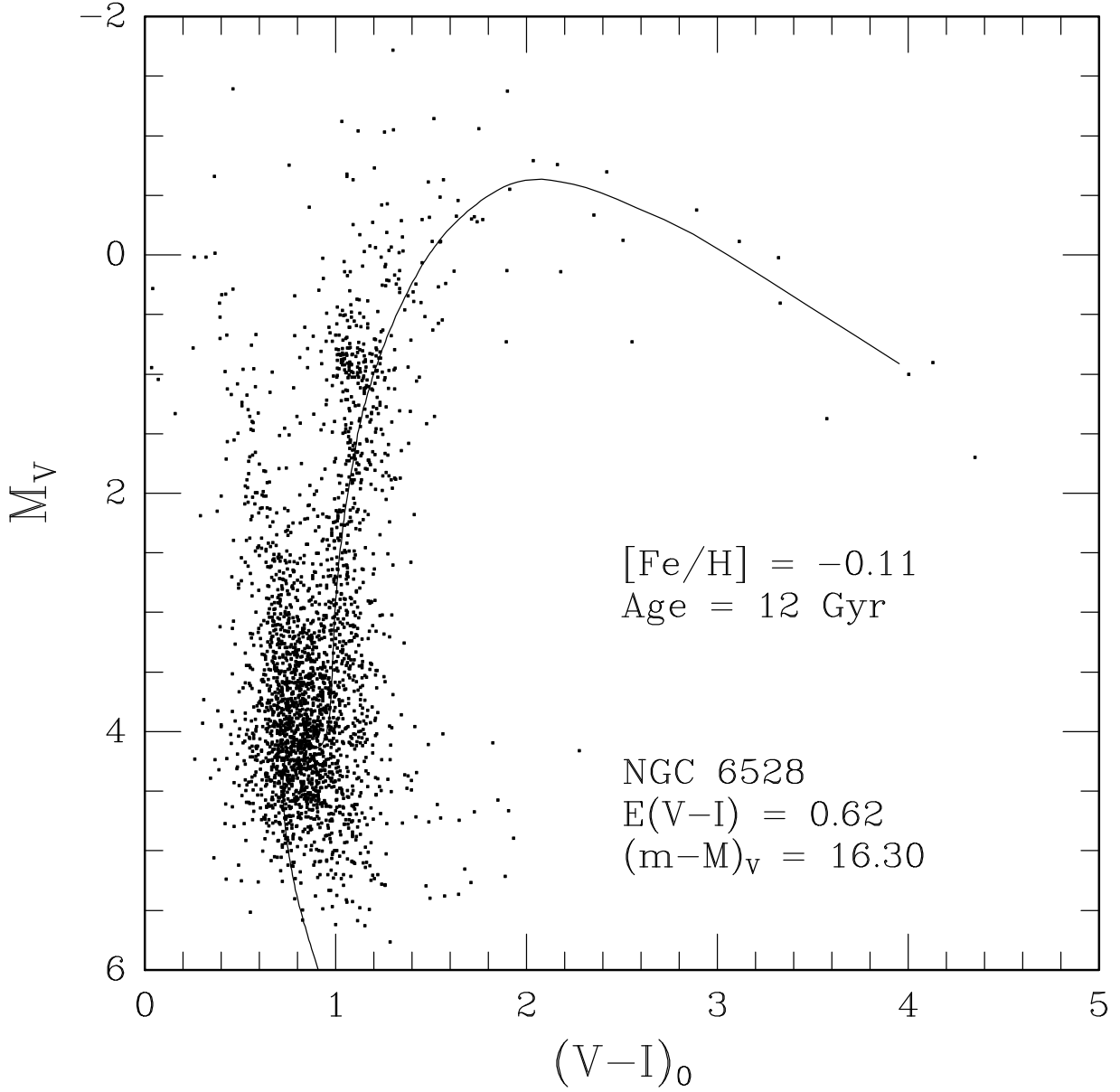}
\caption{Comparison of a 12 Gyr isochrone for [Fe/H] $=-0.11$ and [$\alpha$/Fe]
$=0.0$ with the CMD of NGC$\,$6528 by Richtler et al.~(1998) when the indicated
values of $E(V-I)$ and $(m-M)_V$ are assumed.  The adopted metal abundances are
close to those derived by Zoccali et al.~(2004), who obtained [Fe/H] $=-0.1$
and [$\alpha$/Fe] values between 0.0 (for Ca, Ti) and 0.1 (for O, Mg, Si)
from their spectroscopic study of NGC$\,$6528.}
\label{fig:fig14}
\end{figure}

We conclude this section with another example to show how well our isochrones,
in conjunction with the VC03 color transformations, are able to reproduce
extended RGBs; in this case, that of the metal-rich Bulge globular cluster,
NGC$\,$6528.  To produce the plot shown in Figure~\ref{fig:fig14}, the
observations of NGC$\,$6528 obtained by \citet{rgs98} were tranposed to the
[$(V-I)_0,\,M_V$]--plane in the following way.  First, the observed $V$
magnitudes were converted to absolute $V$ magnitudes by assuming that the
cluster HB clump stars have a mean $M_V \approx 0.9$.  This should be accurate
to within $\pm 0.15$ mag for a GC having [Fe/H] $\approx -0.1$ (\citealt{zbh04})
since, for instance, this is close to the mean $M_V$ of the HB stars in the
younger, but slightly more metal rich, open cluster M$\,$67 (see
Fig.~\ref{fig:fig3}).  Having set the $M_V$ scale, the reddening,
$E(V-I) = 0.62$, then follows from the horizontal shift between the observed
$V-I$ colors and those predicted for the lower RGB portion of an isochrone for
the cluster metallicity; i.e., that part which is fainter than the HB.  (Because
the location of the giant branch is not a very sensitive function of the assumed
age, any isochrone within the range of, say, 8 to 16 Gyr could be used.)

This assumes that the models (for [Fe/H] $\approx 0.0$) are reliable, but there
is ample justification for such an assumption.  For one thing, the predicted
$T_{\rm eff}$ scale has been precisely normalized using the Sun.  For another,
as shown by VC03, their color transformations are in excellent agreement with
the best available empirical color--$T_{\rm eff}$ relations for dwarfs and
giants, as well as with observed color--color diagrams.  (Models for near solar
metallicities are much more secure than those for metal-deficient or
super-metal-rich stars.)

If $E(V-I) = 1.33\,E(B-V)$, as assumed by Zoccali et al (2004; see
\citealt{dwc78}), then our estimate of $E(V-I)$ corresponds to $E(B-V) = 0.47$,
which is in very good agreement with the value adopted by Zoccali et
al.~($0.46\pm 0.04$).  Be that as it may, the main point that we wish to make
here is that the upper RGB segment of the isochrone plotted in
Fig.~\ref{fig:fig14} provides quite a reasonable fit to the cluster
giants all the way to $(V-I)\approx 4.0$.  (As the metallicity is uncertain by
at least $\pm 0.2$ dex, it is easy to obtain an even more extended RGB simply
by assuming a slightly higher metal abundance.  Recall that the $V-I$ colors
and the bolometric corrections relevant to cool stars are extremely sensitive
functions of temperature; see VC03).  Thus, to within the uncertainties of the
many factors at play, our computations do a satisfactory job of reproducing the
giant branches of Galactic GCs over the entire range of their metallicities.

Despite the encouraging agreement of our models with observations (also see
\citealt{vsr00}; BV01, VC03, \citealt{cvg04}), it should be kept in mind that
the uncertainties associated with ``observed" temperatures, chemical 
compositions, and distances, as well as the photometric data themselves (see
\citealt{van05}), permit considerable latitute in the models.  Indeed, coupled
with our poor understanding of, in particular, convection in stellar envelopes
and the physical process (or processes) that are apparently operating to
inhibit diffusion in GC stars, one should not expect, e.g., perfect agreement
between synthetic and observed CMDs.

\section{Conclusions}
\label{sec:conclude}
This investigation presents a total of 72 grids of stellar evolutionary tracks
along with the software needed to generate isochrones for any age within the
ranges encompassed by the tracks, and to predict the numbers of stars along 
them (isochrone probability functions, LFs, CFs).  This includes 60 grids in
which enhancements in the $\alpha$-elements are treated explicitly; i.e., for
each of 20 [Fe/H] values between $-2.31$ and $0.0$, model sequences are provided
for [$\alpha$/Fe] $=0.0, 0.3$, and 0.6.  In each of these cases, isochrones
extending to the RGB tip may be computed for ages from 1 to 18 Gyr.  The
remaining 12 grids are for [Fe/H] values from $-0.60$ to $+0.49$, on the 
assumption of [$\alpha$/Fe] $=0.0$ (in each case) and ranges in mass that are
sufficient to permit the calculation of isochrones down to 0.2 Gyr.

Diffusive processes have not been considered in these computations.  In fact,
as discussed in \S 3, models that take diffusion into account are not able to
explain either the spectroscopic data or the photometric observations of old
star clusters as well as those that neglect this physics.  (As already
mentioned, it is generally thought that there are competing processes at work
in the surface layers of stars that limit the efficiency of diffusion.  However,
our understanding is still in a fairly primitive state.)  The novel feature of
the present models is that convective core overshooting has been treated using
a parameterized form of the Roxburgh criterion, in which the free parameter,
$F_{\rm over}$, is assumed to be a function of both mass and metal abundance.
All of the tests that we have conducted thusfar (including those considered by
P.~D.~Dowler \& D.~A.~VandenBerg, in preparation) support our calibration of
$F_{\rm over}$.  Indeed, our overshooting models are able to reproduce the
turnoff morphologies of cluster CMDs very well, irrespective of their age and
metallicity.  They also satisfy the constraints provided by the binary TZ For,
which has proven to be especially difficult for non-overshooting and
overshooting computations to date (see \citealt{lv02}).  (The data for several
other binaries and open clusters are examined in detail by Dowler \& VandenBerg,
who also compare the present models with overshooting calculations in which
convective cores are extended by some fraction of a pressure scale-height.)
It is our intention to extend the models in the coming year to include, in
particular, the core He-burning phase.

\acknowledgements
We are very grateful to Carme Gallart for providing the observations of
NGC$\,$2155, NGC$\,$2176, and SL$\,$556 that have been used in this study.
Support from the Natural Sciences and Engineering Research Council of
Canada through individual Discovery Grants to D.A.V. and to P.A.B.~is also
acknowledged with gratitude.

\clearpage

\clearpage

\clearpage

\clearpage

\clearpage

\clearpage

\clearpage

\clearpage
 
\clearpage
 
\clearpage

\clearpage

\clearpage

\clearpage

\clearpage

\clearpage
\begin{deluxetable}{rcccccccccc}
\tabletypesize{\footnotesize}
\tablecaption{Initial Chemical Abundances for the VR0A, VR2A, and VR4A Grids 
 of Stellar Models that have [$\alpha$/Fe] $=0.0, 0.3,$ and 0.6, respectively
 \label{tab:tab1}}
\tablewidth{0pt}
\tablehead{ & & & \multispan2\hfil [$\alpha$/Fe] $=0.0$\hfil & &
 \multispan2\hfil [$\alpha$/Fe] $=0.3$\hfil & & \multispan2\hfil [$\alpha$/Fe]
 $=0.6$\hfil \\ \colhead{[Fe/H]} & \colhead{$Y$} & &  \colhead{$Z$} &
 \colhead{File Name} & & \colhead{$Z$} & \colhead{File Name} & & \colhead{$Z$}
 & \colhead{File Name}  } 
\startdata
$-2.31$ & 0.2352 & & 1.000E$-04$ & vr0a-231 & & 1.690E$-04$ & vr2a-231 & &
                   3.070E$-04$ & vr4a-231 \\
$-2.14$ & 0.2353 & & 1.500E$-04$ & vr0a-214 & & 2.540E$-04$ & vr2a-214 & &
                   4.610E$-04$ & vr4a-214 \\
$-2.01$ & 0.2354 & & 2.000E$-04$ & vr0a-201 & & 3.380E$-04$ & vr2a-201 & &
                   6.140E$-04$ & vr4a-201 \\
$-1.84$ & 0.2356 & & 3.000E$-04$ & vr0a-184 & & 5.070E$-04$ & vr2a-184 & &
                   9.210E$-04$ & vr4a-184 \\
$-1.71$ & 0.2358 & & 4.000E$-04$ & vr0a-171 & & 6.760E$-04$ & vr2a-171 & &
                   1.230E$-03$ & vr4a-171 \\
$-1.61$ & 0.2360 & & 5.000E$-04$ & vr0a-161 & & 8.450E$-04$ & vr2a-161 & &
                   1.540E$-03$ & vr4a-161 \\
$-1.53$ & 0.2362 & & 6.000E$-04$ & vr0a-153 & & 1.014E$-03$ & vr2a-153 & &
                   1.840E$-03$ & vr4a-153 \\
$-1.41$ & 0.2366 & & 8.000E$-04$ & vr0a-141 & & 1.352E$-03$ & vr2a-141 & &
                   2.450E$-03$ & vr4a-141 \\
$-1.31$ & 0.2370 & & 1.000E$-03$ & vr0a-131 & & 1.690E$-03$ & vr2a-131 & &
                   3.070E$-03$ & vr4a-131 \\
$-1.14$ & 0.2380 & & 1.500E$-03$ & vr0a-114 & & 2.540E$-03$ & vr2a-114 & &
                   4.610E$-03$ & vr4a-114 \\
$-1.01$ & 0.2390 & & 2.000E$-03$ & vr0a-101 & & 3.380E$-03$ & vr2a-101 & &
                   6.140E$-03$ & vr4a-101 \\
$-0.83$ & 0.2410 & & 3.000E$-03$ & vr0a-083 & & 5.060E$-03$ & vr2a-083 & &
                   9.150E$-03$ & vr4a-083 \\
$-0.71$ & 0.2430 & & 4.000E$-03$ & vr0a-071 & & 6.750E$-03$ & vr2a-071 & &
                   1.220E$-02$ & vr4a-071 \\
$-0.61$ & 0.2450 & & 5.000E$-03$ & vr0a-061 & & 8.430E$-03$ & vr2a-061 & &
                   1.520E$-02$ & vr4a-061 \\
$-0.52$ & 0.2470 & & 6.000E$-03$ & vr0a-052 & & 1.010E$-02$ & vr2a-052 & &
                   1.820E$-02$ & vr4a-052 \\
$-0.40$ & 0.2510 & & 8.000E$-03$ & vr0a-040 & & 1.345E$-02$ & vr2a-040 & &
                   2.410E$-02$ & vr4a-040 \\
$-0.30$ & 0.2550 & & 1.000E$-02$ & vr0a-030 & & 1.676E$-02$ & vr2a-030 & &
                   2.991E$-02$ & vr4a-030 \\
$-0.20$ & 0.2600 & & 1.250E$-02$ & vr0a-020 & & 2.090E$-02$ & vr2a-020 & &
                   3.715E$-02$ & vr4a-020 \\
$-0.11$ & 0.2645 & & 1.500E$-02$ & vr0a-011 & & 2.500E$-02$ & vr2a-011 & &
                   4.425E$-02$ & vr4a-011 \\
$ 0.00$ & 0.2715 & &1.880E$-02$ & vr0a000 & & 3.125E$-02$ & vr2a000 & &
                   5.490E$-02$ & vr4a000 \\
\enddata
\end{deluxetable}

\clearpage
\begin{deluxetable}{rccc}
\tabletypesize{\footnotesize}
\tablecaption{Initial Chemical Abundances for the VRSS Grids of Stellar Models
  that have Scaled-Solar Abundances of the Heavy Elements \label{tab:tab2}}
\tablewidth{0pt}
\tablehead{ \colhead{[Fe/H]} & \colhead{$Y$} & \colhead{$Z$} &
 \colhead{File Name} } 
\startdata
$-0.60$ & 0.24644 & 0.0050 & vrss-060 \\
$-0.52$ & 0.24864 & 0.0060 & vrss-052 \\
$-0.39$ & 0.25304 & 0.0080 & vrss-039 \\
$-0.29$ & 0.25744 & 0.0100 & vrss-029 \\
$-0.19$ & 0.26294 & 0.0125 & vrss-019 \\
$-0.11$ & 0.26844 & 0.0150 & vrss-011 \\
$-0.04$ & 0.27350 & 0.0173 & vrss-004 \\
$+0.00$ & 0.27680 & 0.0188 & vrss000 \\
$+0.13$ & 0.29044 & 0.0250 & vrss013 \\
$+0.23$ & 0.30144 & 0.0300 & vrss023 \\
$+0.37$ & 0.32344 & 0.0400 & vrss037 \\
$+0.49$ & 0.34544 & 0.0500 & vrss049 \\
\enddata
\end{deluxetable}

\clearpage
\begin{deluxetable}{cccc}
\tabletypesize{\footnotesize}
\tablecaption{Globular Cluster [Fe/H] Determinations from Different Sources
 \label{tab:tab3}}
\tablewidth{0pt}
\tablehead{ \colhead{Name} & \colhead{Zinn \& West (1984)} & \colhead{Carretta
\& Gratton (1997)} & \colhead{Kraft \& Ivans (2003)} } 
\startdata
M$\,$15     & $-2.15$ & $-2.12$  & $-2.42$ \\
NGC$\,$6752 & $-1.54$ & $-1.42$  & $-1.57$ \\
NGC$\,$1851 & $-1.36$ & $\ldots$ & $-1.19$ \\
47 Tuc      & $-0.71$ & $-0.70$  & $-0.70$ \\
\enddata
\end{deluxetable}

\end{document}